\documentclass[a4paper,11pt]{article}
\pdfoutput=1 

\usepackage{color}
\usepackage{jinstpub} 
\usepackage{multirow}
\usepackage{array}
\usepackage{todonotes}
\usepackage{graphicx}
\usepackage{rotating}
\usepackage{makecell, multirow}
\usepackage{booktabs}
\usepackage{url}
\usepackage{caption}
\usepackage{subcaption}
\usepackage{lineno}
\usepackage{textcomp}
\title{Enhancing Neutrino Event Reconstruction with Pixel-Based 3D Readout for Liquid Argon Time Projection Chambers}

\author[a]{C.~Adams}
\author[b,c]{M.~Del Tutto}
\author[d]{J.~Asaadi }
\author[c]{M.~Bernstein}
\author[e]{E.~Church}
\author[c]{R.~Guenette}
\author[c,1]{J. M. Rojas \note{Currently at ICTP-SAIFR/IFT-UNESP, S\~ao Paulo, Brazil}}
\author[d]{H.~Sullivan}
\author[d]{A.~Tripathi }

\affiliation[a]{Argonne National Laboratory, Lemont, IL}
\affiliation[b]{Fermi National Accelerator Laboratory, Batavia, IL}
\affiliation[c]{Harvard University, Cambridge, MA}
\affiliation[d]{University of Texas-Arlington, Arlington, TX}
\affiliation[e]{Pacific Northwest National Laboratory, Richland, WA}

\emailAdd{corey.adams@anl.gov}

\abstract{In this paper we explore the potential improvements in neutrino event reconstruction that a 3D pixelated readout could offer over a 2D projective wire readout for liquid argon time projection chambers. We simulate and study events in two generic, idealized detector configurations for these two designs, classifying events in each sample with deep convolutional neural networks to compare the best 2D results to the best 3D results. In almost all cases we find that the 3D readout provides better reconstruction efficiency and purity than the 2D projective wire readout, with the advantages of 3D being particularly evident in more complex topologies, such as electron neutrino charged current events. We conclude that the use of a 3D pixelated detector could significantly enhance the reach and impact of future liquid argon TPC experiments physics program, such as DUNE.}


\begin{document}
\maketitle
\flushbottom

\newcommand{\STAB}[1]{\begin{tabular}{@{}c@{}}#1\end{tabular}}

\section{Introduction}\label{Sec:intro}
Large-scale noble element time projection chambers (TPCs) play a central role in current and future high energy physics experiments \cite{LZ, xenon1t, pandax, next, darkside, exo, argoneut, Acciarri:2019wgd,uboone,sbn,dune}. As charged particles traverse the bulk material they produce ionization electrons and scintillation photons. An external electric field drifts the ionization electrons towards the anode of the detector, where they are collected on charge sensitive readout. The combined measurement of the scintillation light, providing the $t_0$ of the event, and the arrival time of the ionization charge at the anode allows for a 3D reconstruction of the original charged particle topology. Thus the TPC provides a fully active tracking detector with calorimetric reconstruction capabilities without instrumenting the bulk volume of the detector.

\begin{figure}
    \centering
    \includegraphics[scale = 0.4]{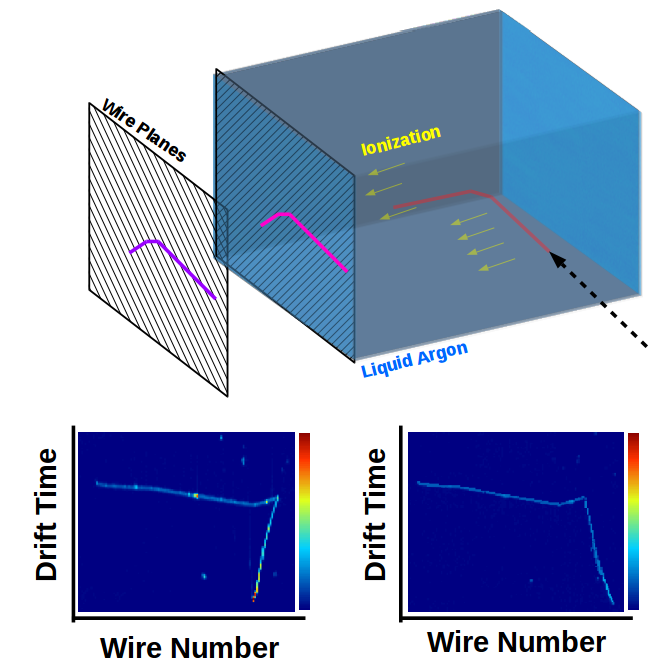}
    \caption{Top: Schematic of the LArTPC principle. Incoming particles (black dashed arrow) enter the TPC and ionize the argon atoms along its path (red line). These ionization electrons (yellow arrows) are drifted towards the readout planes, where they induce signal in the innermost (induction) plane(s) (pink line) and are collected on the last (collection) plane (purple line). Bottom: Two event displays from an event interaction in a TPC. The 2D projection for each plane, where the signal on each wire is shown in function of its drift time since $t_0$. Image source: \cite{Acciarri:2019wgd}.}
    \label{fig:LArTPCExample}
\end{figure}

Liquid argon time projections chambers (LArTPCs) have become a commonly used technology for neutrino physics \cite{argoneut,Acciarri:2019wgd, uboone,sbn,dune}. The standard method for reading out the ionization charge in a LArTPC utilizes consecutive planes of charge sensing wires, as illustrated in Fig.~\ref{fig:LArTPCExample}, to measure two of the three space coordinates. The third spatial coordinate is measured using the arrival time of the charge on the wires relative to the time of the scintillation light ($t_0$). Each plane of charge-sensitive wires generates a two-dimensional image (wire number versus time), also illustrated in Fig.~\ref{fig:LArTPCExample}. Analyzing the patterns in the multiple images generated by the consecutive wire planes and matching them across the wire planes allows for the 3D reconstruction of the interaction. This concept of projective charge readout for LArTPC has considerable legacy experience in the community, but has an intrinsic limitation in resolving ambiguities which arise from ionization depositions which travel parallel or perpendicular to the wire planes.  Additionally, complex topologies (such as electromagnetic showers or deep inelastic scattering events) can be degraded by projective readouts  with overlapping projections, where no such overlap occurs in a 3D readout. Such inherent pathologies make the event reconstruction very challenging in some cases. Future event reconstruction techniques may overcome these challenges, for example \cite{wirecell}, however detector instrumentation approaches that do not have this intrinsic challenge offer a more direct solution.

\begin{figure}
    \centering
    \includegraphics[scale = 0.35]{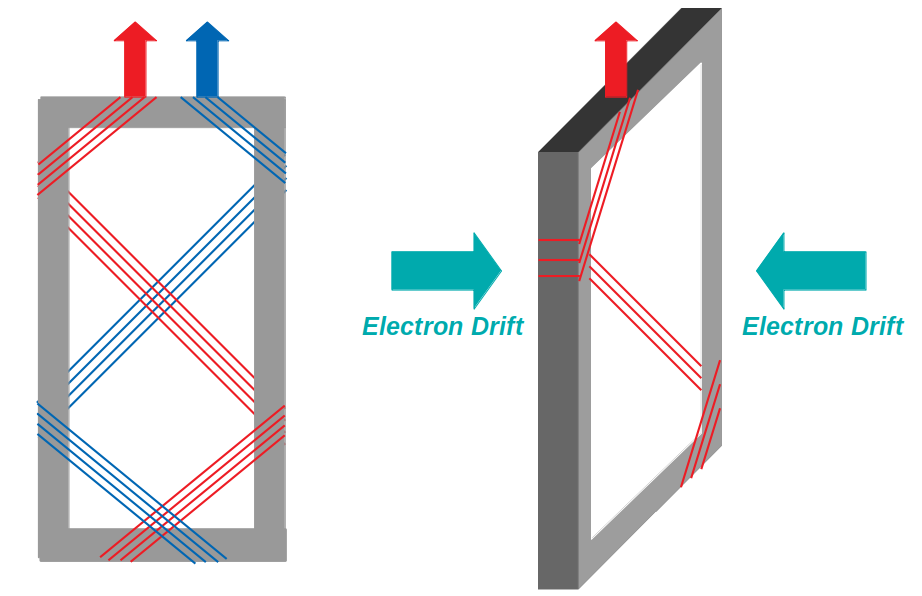}
    \caption{Illustration of the wrapped wire configuration used in kiloton scale LArTPCs. The wires from two inductions planes are visible in red and blue, and they wrapped around the grey support structure. Left: Front view of the wrapped wire design. Right: Side view of the design where the two drift volumes on either side of the structure are readout by the same wires. }
    \label{fig:WrappedWire}
\end{figure}

In order to reduce the total number of wires which have to be read out in kiloton scale LArTPCs, such as the Deep Underground Neutrino Experiment (DUNE) detectors \cite{dune_tdr_1,dune_tdr_2,dune_tdr_3}, the design of the readout anode plane assembly (APA) will have the induction planes wires wrapped around the support structure (as illustrated in Fig.~\ref{fig:WrappedWire}). A single APA  will read out two drift volumes on either side of the structure. While the wrapped wire geometry significantly reduces the number of channels which must be read out, it introduces even more potential ambiguities. These ambiguities can be an issue in particular for high energy events with significant activity in the detector, such as electromagnetic activity coming from electron neutrino charged current interactions. This is potentially problematic since these events are the primary signal in long-baseline neutrino oscillation experiments. A non-projective readout could directly mitigate these issues.

Pixelated readout schemes were not considered viable for LArTPCs until recently because of the larger number of readout channels and the power dissipation requirements at cryogenic temperatures for existing LArTPC readout technologies. The number of pixels needed to equate the wire spatial resolution can be two or three orders of magnitude higher, with a similar increase of the number of electronics readout channels, data rates and power dissipation. This makes such a solution untenable except for very small detectors. A truly transformative step forward for future LArTPCs is the ability to build a fully pixelated and low-power charge readout. The significant advantage of a low-power, pixel-based charge readout for use in LArTPCs has independently inspired two complementary research approaches to address this. The LArPix \cite{larpix} and Q-Pix \cite{Nygren:2018rbl} solutions are currently under development.

In this paper, we study the potential improvements a 3D pixel-based readout provides over the 2D projective readout. We compare event identification of both approaches and quantify the potential gain. To mitigate the intrinsic differences in the reconstruction quality for 2D and 3D detectors for neutrino interactions, we use techniques based on machine learning and convolutional neural networks.  We train all networks from random initialization on the same dataset using the same neutrino events, and compare the best 2D results to the best 3D results. Based on recent results from computer vision research \cite{alexnet, resnet, inception_net} and its applications to neutrino physics \cite{uboone_ml} we believe that this is the fairest way to compare 2D and 3D event reconstruction.

Section \ref{Sec:tools} provides an overview of the methods used to simulate neutrino interactions and to project them to either the 2D wire readout or 3D pixel  readout. Section \ref{Sec:DL} describes the implementation of the deep convolutional neural networks for comparison between 2D and 3D readouts. Section \ref{Sec:2dvs3d} presents the results based on the various typologies and performances of the classifications. 

\section{Simulation of neutrino events for 2D and 3D readouts}\label{Sec:tools}
We simulate neutrino interactions inside a block of liquid argon of $3.6 \times 2.0 \times 5.0$ m$^3$ in $x$, $y$ and $z$ respectively, utilizing the \texttt{LArSoft} event simulation framework \cite{larsoft} and a customized pixel simulation module \cite{pixgen:2019}. The chosen simulation volume represents approximately 50 tons of liquid argon, the approximate volume that would be seen by one APA wire structure in the current design of DUNE \cite{dune}.

We simulate neutrino interactions using the \texttt{GENIE} neutrino event generator \cite{genie}. The corresponding flux of neutrinos as a function of energy is chosen to match that for the DUNE experiment far detectors \cite{dune_flux}. The particles created from the neutrino interactions are propagated through the liquid argon volume with the \texttt{GEANT4} simulation package \cite{geant}. 

\begin{figure}
    \centering
    \includegraphics[angle=270,width=0.5\linewidth]{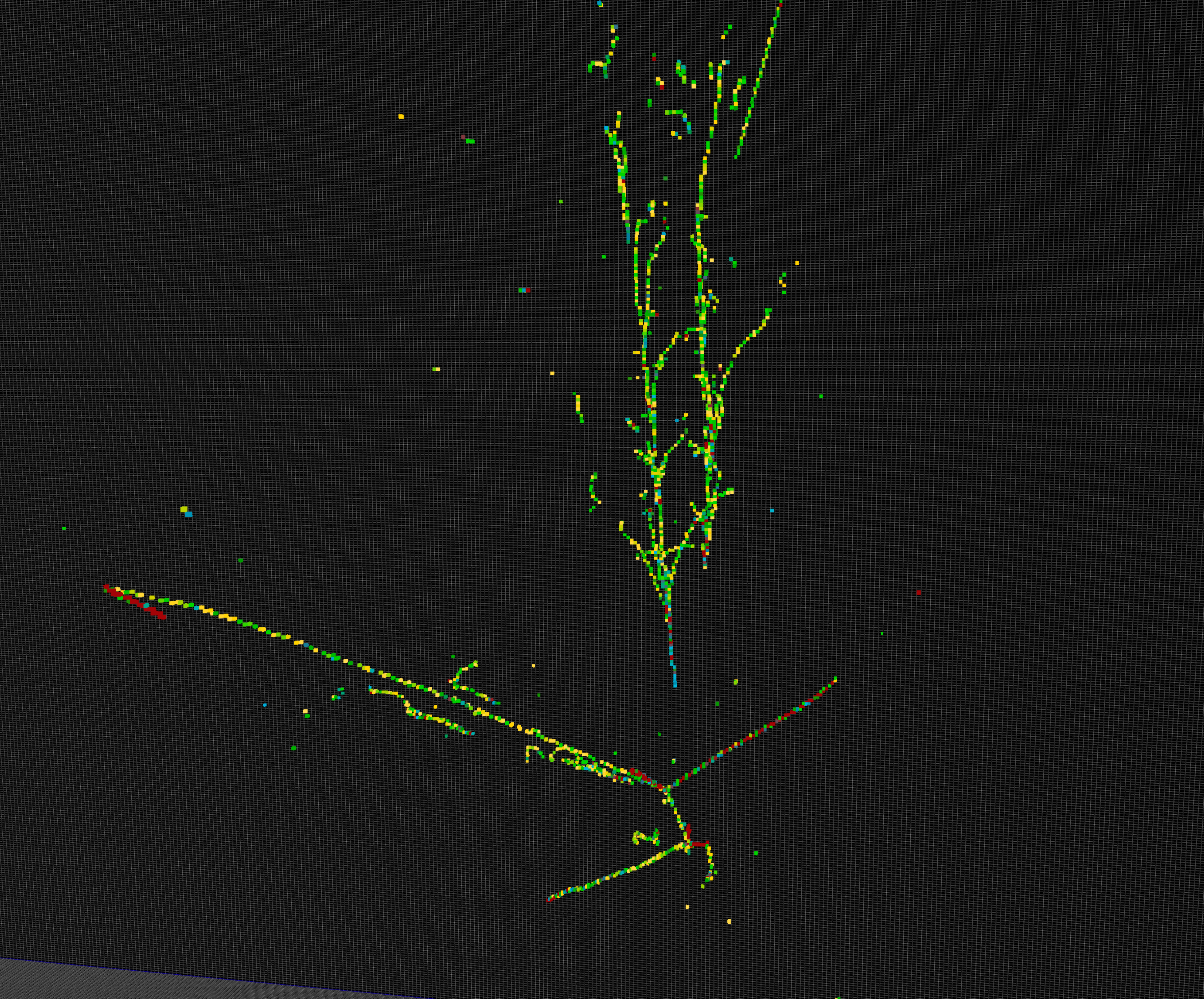}
    \caption{A view of a 3D simulated neutrino interaction with pixel size of 4mm$^{3}$.}
    \label{fig:3dVoxel}
\end{figure}

For each neutrino interaction or event, we record the location of each energy deposition simulated by \texttt{GEANT4}. The detector is then segmented into 4 mm$^3$ 3D elements which directly represent the 3D pixel readout images. If two or more energy depositions are found in a single 3D pixel, the energy is summed. Figure~\ref{fig:3dVoxel} shows an example of an event simulated in the 4 mm$^3$ environment for the 3D readout. 

\begin{figure}
    \centering
    \includegraphics[width=0.3\linewidth]{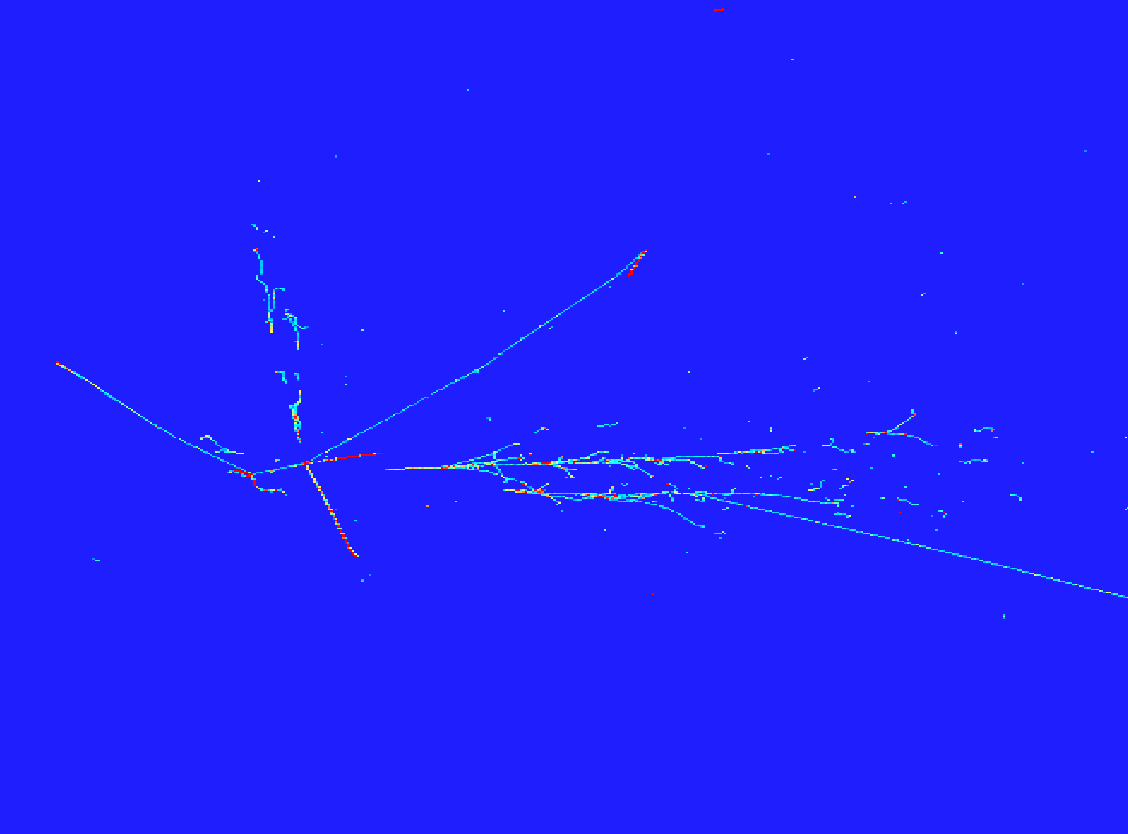}
    \includegraphics[width=0.3\linewidth]{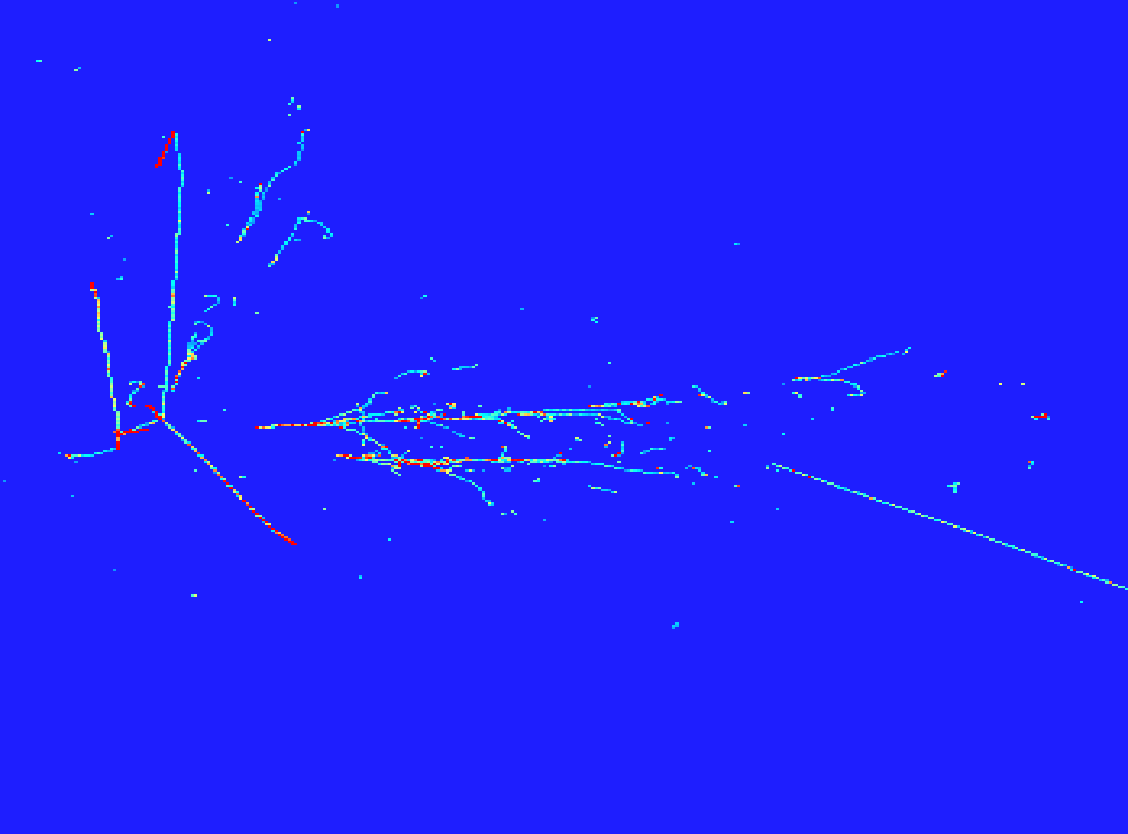}
    \includegraphics[width=0.3\linewidth]{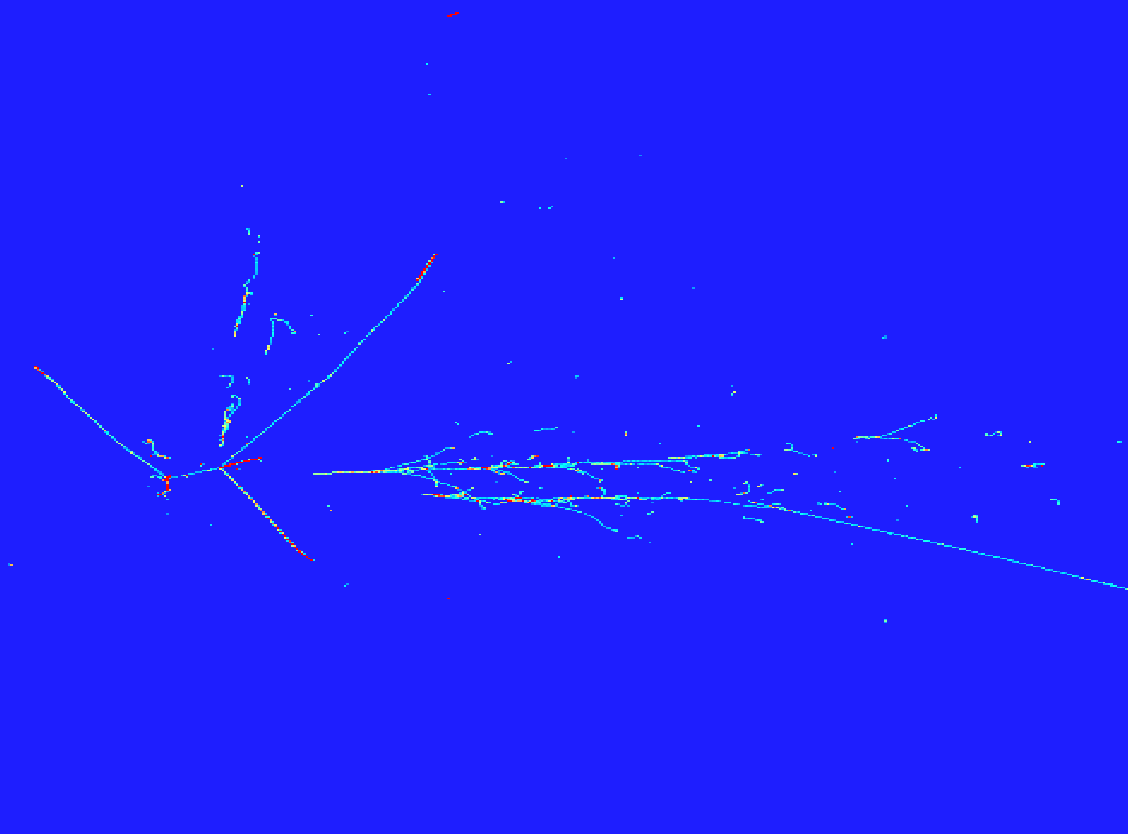}
    \caption{The three 2D projections of the neutrino interaction pictured in Fig.~\ref{fig:3dVoxel}. The
    X-axes represent the wire numbers for each wire planes and the Y-axes the drift time coordinate. }
    \label{fig:2DProjection}
\end{figure}

To image the 2D wire readout, the \texttt{GEANT4} energy depositions are projected onto three wire planes with 2D pixel sizes of 4 mm$^2$. We use wire planes orientation of angles $\theta$ equal to -35.7$^{\circ}$, +35.7$^{\circ}$, and 0$^{\circ}$ to match the DUNE geometry for the two induction planes and the collection plane respectively. This is done via a rotation in 3D followed by a projection of the 3D pixels to 2D in the following way:
\begin{alignat*}{2}
     x_{2D} &= \cos(\theta)\,z_{3D} + \sin(\theta)\,y_{3D} \,\, &&(\text{Wire Projection}), \\ 
     y_{2D} &= x_{3D}                                           &&(\text{Drift Distance}),  \\
\end{alignat*}
where $\theta$ is each of the angles described above.  For each projection, the vertical coordinate $y_{2D}$ is set to be exactly the drift distance $x_{3D}$.  The horiztonal projection coordinate $x_{2D}$ is the rotation of the vertical ($y_{3D}$) and beam ($z_{3D}$) coordinates.  Fig.\ref{fig:2DProjection} shows the 2D projection of the same neutrino event depicted in Fig.\ref{fig:3dVoxel} for each of the three protective views. As in the 3D case, if multiple charge depositions are in the same 2D pixel (either through inherent resolution pixelization or by the lossy nature of the projection), they are summed.

All this information is stored in \texttt{larcv3} \texttt{hdf5} file format \cite{larcv3} along with associated Monte Carlo generator information and is used for analysis.

We do not simulate any detector-level effects and we assume that both the 2D projective and 3D pixel readouts perform perfectly. Effects such as recombination, diffusion, and corrections for electron lifetime are expected to be similar for both readout technologies and thus do not warrant comparison at this point. More complicated effects, such as electronic noise, dynamic range, and deconvolution of electronics response, which are specific to a particular choice of readout technology, are not taken into account in our studies. One important effect which is not incorporated into this work is the effect which arises when particles travel parallel to a series of wires in the 2D projective readout \cite{ub_sig_proc_1}. Events with this topology are particularly challenging to reconstruct using 2D projective readout  \cite{wirecell}. Similarly, the effect of particles traveling perpendicularly to the 2D or 3D readout and the distortion of the signal shapes are also omitted. The absence of such detailed detector effects allows us to compare an idealized 3D readout to an idealized version of the 2D projective readout, and any additional complication will be studied in further work.

We generated a sample of 10M neutrino events on the Theta computing system, a Cray XC40 Supercomputer hosted at Argonne Leadership Computing Facility, using Balsam, an HPC workflow manager \cite{balsam}. For training of the networks, we desire to have equal parts of the relevant neutrino interaction rates, each with an energy spectrum as expected from the DUNE far detector flux \cite{dune_flux}.  The train, test and validation datasets are equal parts of the following samples:
\begin{enumerate}
    \item \textbf{Electron Neutrino ($\nu_e$) Charged Current (CC) Events}, where only electron (anti)neutrinos from the DUNE flux are simulated and interacted via charged current processes.
    \item \textbf{Muon Neutrino ($\nu_{\mu}$) Charged Current (CC) Events}, where only muon (anti)neutrinos from the DUNE flux are simulated and interacted via charged current processes.
    \item \textbf{Neutral Current (NC) Events}, where all (anti)neutrino flavors from the DUNE flux are simulated, but only interacted via neutral current processes.
\end{enumerate}
We simulated approximately 2.5M events in each of the above 3 categories, which are merged into a random order in the training datasets.  Additionally, we simulated 2.5M events without any restrictions on the \texttt{GENIE} event generator or flux, so that we may run inference on this dataset and simulate a DUNE analysis based on the output of the trained neural networks.  The datasets used in this paper, both the training, testing and validation as well as the beam simulation, totalling approximately 1TB, are freely available at \cite{osf_samples}.

\section{Comparison methodology and network training}\label{Sec:DL}
One of the main goals for noble element neutrino detectors is to correctly reconstruct and select, for every interaction and with high efficiency, the type of neutrino interaction. To approach this problem we use deep convolutional neural networks (CNNs) for both 2D projective and 3D pixel-based readout. CNNs are a particular family of Deep Learning (DL) networks which are part of a rich area of active research beyond the scope of this paper; some example resources can be found here ~\cite{ml_neutrinos_nature,uboone_ml,goodfellow_book}.


%

In order to compare the physics sensitivity of 2D projective and 3D pixel-based LArTPCs, we choose to focus on the classification tasks rather than developing a full CNN-based reconstruction. The correct classification of events is one of the most important components of having a full event reconstruction which can achieve both high efficiency and low background contamination, as needed by experiments like DUNE \cite{dune_tdr_1, dune_tdr_2, dune_tdr_3}. Developing the full reconstruction chain is a compelling and long-term approach which will be addressed by future work and is hopefully enabled by making the dataset and software used here open source. The ultimate physics sensitivity of any experiment, regardless of detector technology, is driven significantly by the success or failure of a full reconstruction chain which requires a large team of collaborators. We do not attempt to emulate such a full reconstruction chain nor comment on the full physics sensitivity of these experiments. Instead, the ability to classify each topology serves as a surrogate for reconstruction, with the hypothesis that a well trained neural network will extract as much information as possible from either 2D or 3D images. This allows for adequate and fair comparison between the two detector readout technologies without having a full reconstruction process for either.

One key challenge in the application of these networks to the tasks of neutrino event classification is that, opposed to traditional image processing, the events are quite sparse. The 2D images have, for example, an average pixel occupancy of $\sim 0.09\%$ for the 3 projections and the 3D pixel occupancy is lower, averaging $\sim 3 \times 10^{-4}$\%. Even the densest images in the dataset, which are 2D, have just 1\% of their pixels with non-zero values. Additionally, the full resolution images are quite large in size, reaching $\mathcal{O}$(0.5M) pixels in 2D and $\mathcal{O}$(0.5B) pixels in 3D. Processing the dense 2D images in a convolutional network is challenging on current computing hardware; processing the dense 3D images is a technical impossibility. We adopt a technique of sparse convolutional neural networks, pioneered in \cite{SubmanifoldSparseConvNet} and demonstrated to effectively work on some neutrino physics sparse images \cite{laura_D_paper}.  We also employ distributed learning techniques in order to accelerate the training time. All results shown here are trained on Summit~\cite{summit-website}, currently the fastest high performance computing (HPC) system in the world, at Oak Ridge National Lab. We use \texttt{pytorch} \cite{pytorch} to implement our networks and \texttt{Horovod} software \cite{horovod} to perform data parallel network training. Each Summit node has 6 NVIDIA V100 GPUs. We use a batch size of 256 images per GPU, for a single-node batch size of 1536 images. All of the software for training, inference and analysis is open source and available in~\cite{our_github}.

For this work the networks are trained to perform a simultaneous classification of 4 categories, which in total represent a large set of the interesting final state configurations for physics analyses:
\begin{enumerate}
    \item \textbf{Neutrino Interaction:} Each event is classified to be either $\nu_e$ CC, $\nu_\mu$ CC, or NC.
    \item \textbf{Proton Multiplicity:} Each event is classified to have 0, 1, or $\geq$ 2 protons present in the interaction. The truth label for the number of protons from the neutrino interaction requires the final state proton to have a kinetic energy threshold of~$\geq 50$~MeV.
    \item \textbf{Charge Pion Presence:} The event is classified by whether or not there are charged pions ($\pi^{\pm}$) present in the neutrino interaction with a kinetic energy threshold of~$\geq 50$~MeV.
    \item \textbf{Neutral Pion Presence:} The event is classified by whether or not there are neutral pions ($\pi^0$) present in the neutrino interaction with a kinetic energy threshold of~$\geq 50$~MeV.
\end{enumerate}

\begin{figure*}
\includegraphics[width=1.0\textwidth]{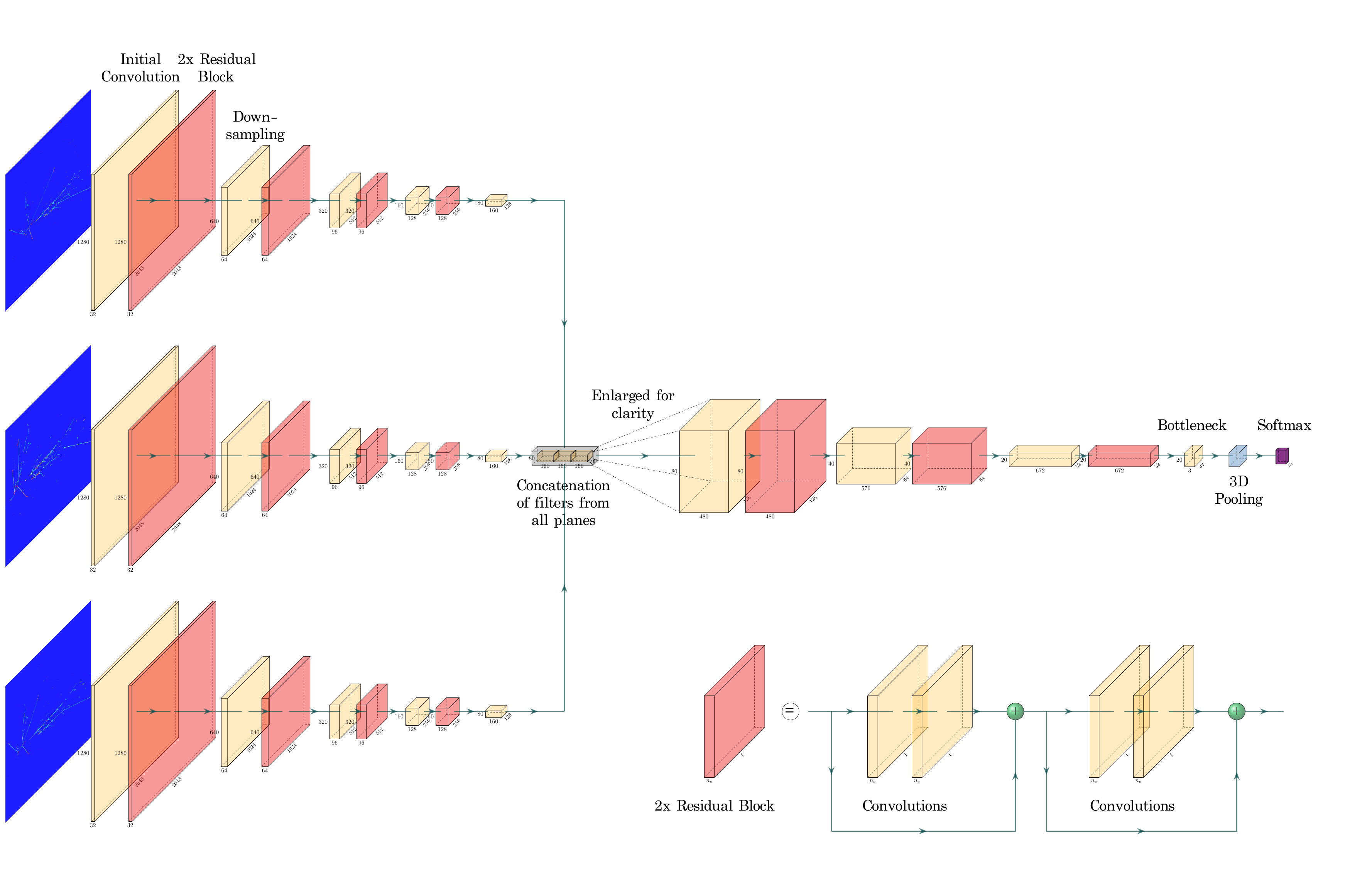}
\caption{Network architecture for classification of 2D images. Each plane is first evaluated separately for several convolutions and downsamples; the planes are then concatenated together to apply more convolutions. Residual blocks~\cite{resnet} are used throughout the network. The input images are 2D sparse images made with~\cite{plotneuralnet}.}
\label{fig:architecture_2d}
\end{figure*}

Fig.~\ref{fig:architecture_2d} shows our network architecture for the 2D projective readout networks, which we describe here.  A ``Siamese Tower'' style network is utilized where each of the three projected images is evaluated separately for several convolutional and downsampling layers.  After a time, the intermediate activations are concatenated together and more convolutions are applied to all planes together. The final layer is used as a learned feature representation for each of the above categories.  Short, independent paths after the feature representation are used for each of the four classification categories and the total loss is the softmax cross entropy of each individual category summed together.

\begin{figure*}
\includegraphics[width=1.0\textwidth]{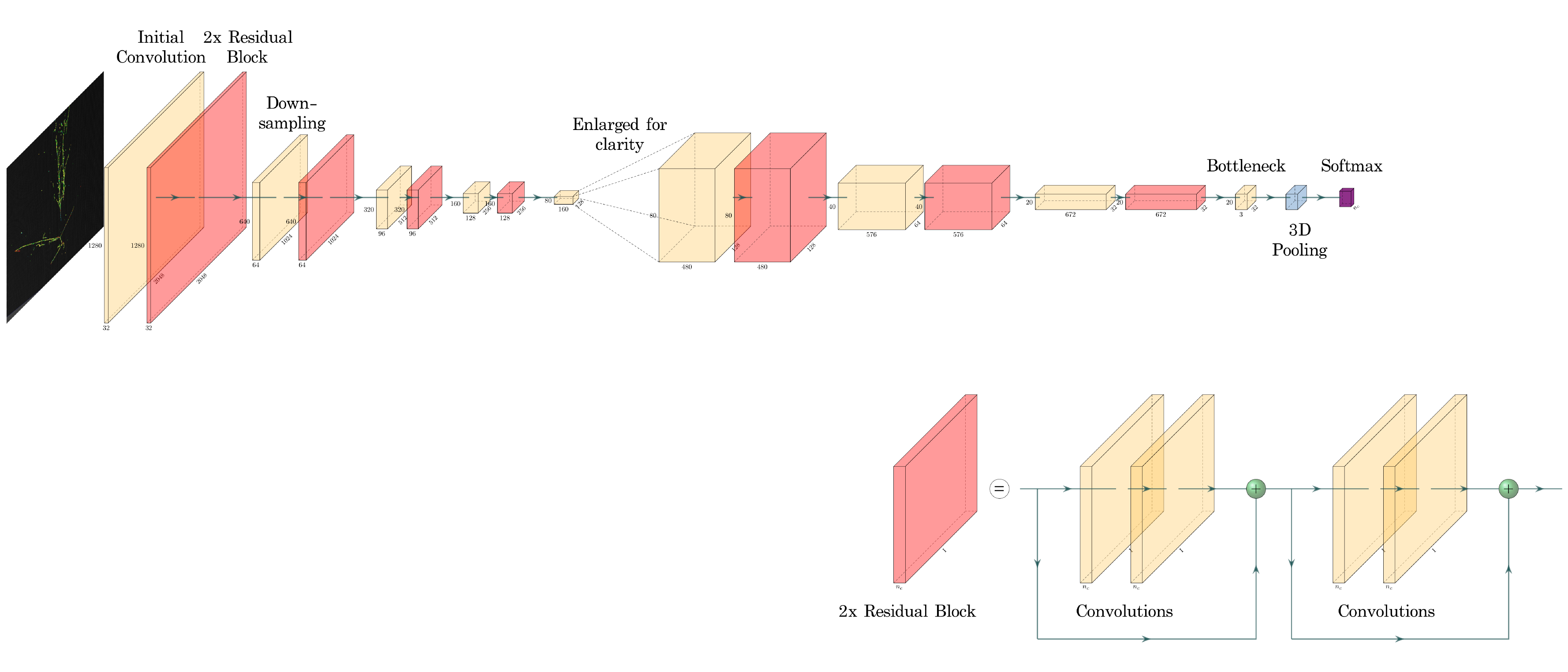}
\caption{Network architecture for classification of 3D images. Similar to the 2D architecture shown in Fig.~\ref{fig:architecture_2d}, but here a single path deep neural network is used. Although the representation shows a 2-dimensional image for clarity, the actual network works with a 3D sparse image made with~\cite{plotneuralnet}.}
\label{fig:architecture_3d}
\end{figure*}

In 3D, we use exactly the same loss functions as in 2D, but use a single path deep neural network instead of a ``Siamese Tower'' style. Fig.~\ref{fig:architecture_3d} shows the structure of this network. In this way, the 2D and 3D networks are as similar as possible. This desire to use the 3D generalization of the 2D network is a further effort to place the 2D and 3D reconstructions on a level playing field. 

\subsection{Network training}\label{Sec:training}

Due to the differences in the two networks, including dataset dimensionality, kernel shapes (3D kernel is $3 \times 3$ while 2D kernel is $3 \times 3 \times 3$), and number of overall parameters, we do not necessarily expect to get the same performance for networks when using the same batch size and other hyperparameters.  Therefore, we explore several different hyperparameters and make a comparison of the best 2D network performance to the best 3D network performance.  A full hyperparameter exploration could be open-ended and the proper exploration of the space of 2D and 3D networks will require a full optimization of hyperparameters via a dedicated analysis, such as with DeepHyper \cite{deephyper1,deephyper2}.  Such a hyperparameter scan is beyond the scope of this work, but is anticipated as a future result exploring this space more completely.

\begin{figure}
\centering
    \begin{subfigure}[]{0.495\textwidth}
        \includegraphics[width=\textwidth]{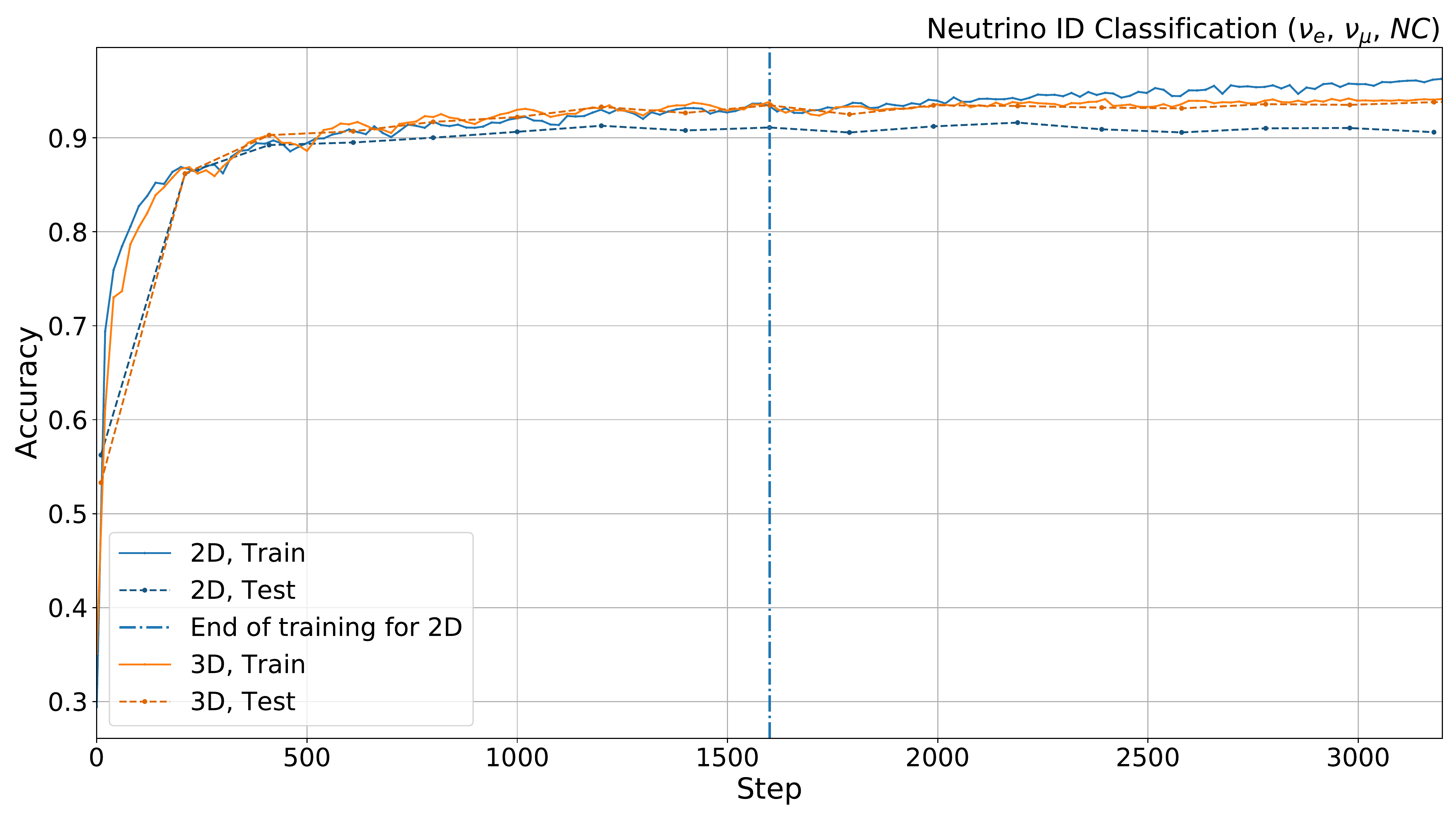}
        \caption{}
        \label{fig:accuracy_neut}
    \end{subfigure}
    \begin{subfigure}[]{0.495\textwidth}
        \includegraphics[width=\textwidth]{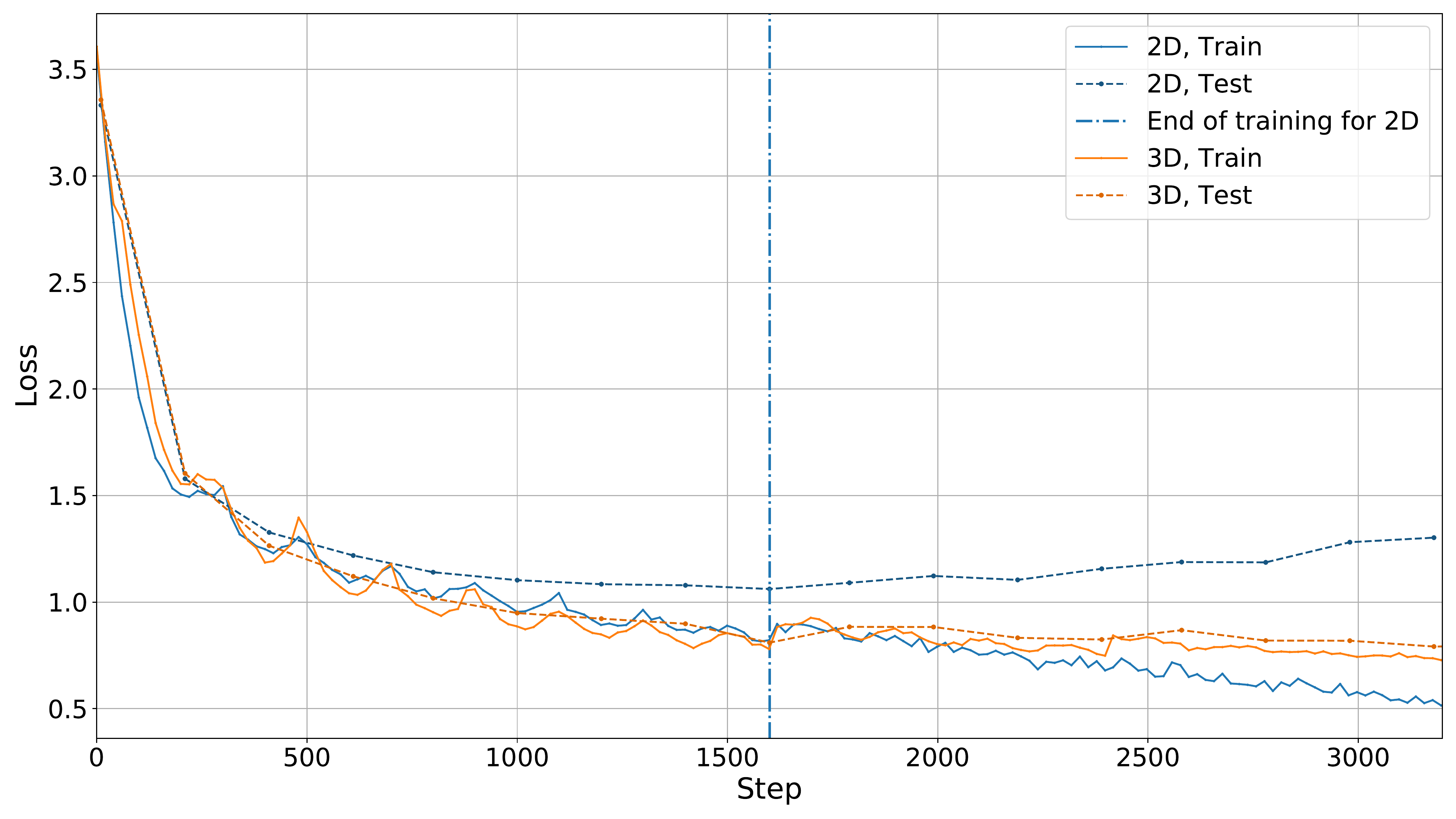}
        \caption{}
        \label{fig:loss}
    \end{subfigure}
    \caption{
        Training and testing accuracy (\subref{fig:accuracy_neut}) and loss (\subref{fig:loss}) for both the 2D (blue curves) and 3D (orange curves) networks.  Shown here is just the neutrino classification accuracy. Although the 2D curves are shown up to iteration step 3200, for the following studies we pick the trained model as obtained at iteration 1600 (vertical dashed blue line), before the network begins overfitting.
    }
\label{fig:accuracy_loss}
\end{figure}

Figure~\ref{fig:accuracy_loss} compares the training for the 2D and 3D networks in our best runs for each network.  The testing accuracy is used to monitor for over-fitting during training, and the network with the highest testing accuracy at the end of the training in both 2D and 3D is used for the comparisons in the subsequent sections. We show on the figure the total loss for each network and the accuracy of neutrino classification. The accuracies for the other three categories are available in Appendix~\ref{app:plots}.

\section{Results of the 2D vs 3D analysis}\label{Sec:2dvs3d}

We present here several quantitative metrics for comparison of 2D and 3D performances.  First, we compare the correctness of the networks on the four different classification categories and present the confusion matrices for both 2D and 3D.  Additionally, the DUNE neutrino flux will have very few electron neutrinos compared to muon neutrinos, and so the raw classification accuracies do not tell the whole story.  Therefore we present several toy analyses making selections of events from the beam simulation dataset, as well as comparing 2D vs. 3D selection efficiencies and purities.




Table~\ref{tab:accuracies} provides a high-level summary of the accuracy obtained in the validation data set for the best 2D and 3D networks. The 3D network outperforms the 2D network in every task of accurately classifying the various categories. Tables~\ref{tab:confusion_matrix_nu},~\ref{tab:confusion_matrix_p},~\ref{tab:confusion_matrix_cpi} and~\ref{tab:confusion_matrix_npi} show the rates at which the true labels (shown horizontally) for the various sub-categories of neutrino events are classified versus the predicted labels (shown vertically) from the network.  The diagonals in each table represent the per-category accuracy and the off-diagonal elements show the mislabeling each category has for the two readouts. 

For all accuracies the 3D pixel-based readout performs better than the 2D readout except for predicting the absence of neutral pions ($\pi^0$) in the event, where it performs at parity with the 2D network. The mislabeling of the various categories is also better for the 3D network for nearly all categories except in the mislabeling of events with one proton and events with no protons. Here the 3D network under-performs at the level of $\sim 1\%$.  

\begin{table}
\centering
\caption{Accuracy in the validation set for the best 2D and the best 3D networks to correctly classify neutrino events.}
\label{tab:accuracies}
\begin{tabular}{c c c}
\toprule
                        & \multicolumn{2}{c}{Accuracy  [\%]} \\
Category                &  3D & 2D   \\
\midrule
Neutrino Interaction    &  \bfseries 94 & 91   \\
Proton Multiplicity     &  \bfseries 91 & 87   \\
Charge Pion Presence    &  \bfseries 94 & 91   \\
Neutral Pion Presence   &  \bfseries 95 & 94   \\
\bottomrule
\end{tabular}
\end{table}

\begin{table}
    \centering
    \caption{Confusion matrix for the neutrino interaction classification.}
    \begin{tabular}{c c | c c c c|c c c c}
        \toprule
        & & \multicolumn{3}{c}{3D}          &&& \multicolumn{3}{c}{2D} \\
        & & \multicolumn{3}{c}{Truth Label} &&& \multicolumn{3}{c}{Truth Label} \\
        & & $\nu_e$ CC & $\nu_\mu$ CC & NC  &&& $\nu_e$ CC & $\nu_\mu$ CC & NC     \\
        \midrule
        \multirow{3}{*}{\STAB{\rotatebox[origin=c]{90}{Pred.}}}
        \multirow{3}{*}{\STAB{\rotatebox[origin=c]{90}{Label}}}
        & $\nu_e$ CC      & \bfseries 0.959 & 0.013           & 0.019           &&& 0.928 & 0.019 & 0.025  \\
        & $\nu_\mu$ CC    & 0.018           & \bfseries 0.952 & 0.072           &&& 0.020 & 0.908 & 0.070  \\
        & NC              & 0.023           & 0.035           & \bfseries 0.908 &&& 0.052 & 0.073 & 0.904  \\
        \bottomrule
    \end{tabular}
    \label{tab:confusion_matrix_nu}
\end{table}

\begin{table}
    \centering
    \caption{Confusion matrix for the proton multiplicity classification.}
    \begin{tabular}{c c | c c c c|c c c c}
        \toprule
        & & \multicolumn{3}{c}{3D}          &&& \multicolumn{3}{c}{2D} \\
        & & \multicolumn{3}{c}{Truth Label} &&& \multicolumn{3}{c}{Truth Label} \\
        & & $N_{p} = 0$ & $N_{p} = 1$ & $N_{p} \geq 2$  &&& $N_{p} = 0$ & $N_{p} = 1$ & $N_{p} \geq 2$     \\
        \midrule
        \multirow{3}{*}{\STAB{\rotatebox[origin=c]{90}{Pred.}}}
        \multirow{3}{*}{\STAB{\rotatebox[origin=c]{90}{Label}}}
        & $N_{p} = 0$      & \bfseries 0.928 &  0.076           & 0.005            &&& 0.841 &  0.064 & 0.005  \\
        & $N_{p} = 1$      & 0.062           &  \bfseries 0.884 & 0.059            &&& 0.143 &  0.853 & 0.069  \\
        & $N_{p} \geq 2$   & 0.010           &  0.040           & \bfseries 0.936  &&& 0.016 &  0.084 & 0.926  \\
        \bottomrule
    \end{tabular}
    \label{tab:confusion_matrix_p}
\end{table}

\begin{table}
    \centering
    \caption{Confusion matrix for the charged pion presence classification.}
    \begin{tabular}{c c | c c c|c c c}
        \toprule
        & & \multicolumn{2}{c}{3D}          &&& \multicolumn{2}{c}{2D} \\
        & & \multicolumn{2}{c}{Truth Label} &&& \multicolumn{2}{c}{Truth Label} \\
        & & $N_{\pi^{\pm}} = 0$ & $N_{\pi^{\pm}} \geq 1$ &&& $N_{\pi^{\pm}} = 0$ & $N_{\pi^{\pm}} \geq 1$     \\
        \midrule
        \multirow{2}{*}{\STAB{\rotatebox[origin=c]{90}{Pred.}}}
        \multirow{2}{*}{\STAB{\rotatebox[origin=c]{90}{Label}}}
        & $N_{\pi^{\pm}} = 0$      & \bfseries 0.934 &  0.052            &&& 0.923 &  0.101 \\
        & $N_{\pi^{\pm}} \geq 1$   & 0.066           &  \bfseries 0.948  &&& 0.077 &  0.899 \\
        \bottomrule
    \end{tabular}
    \label{tab:confusion_matrix_cpi}
\end{table}

\begin{table}
    \centering
    \caption{Confusion matrix for the neutral pion presence classification.}
    \begin{tabular}{c c | c c c|c c c}
        \toprule
        & & \multicolumn{2}{c}{3D}          &&& \multicolumn{2}{c}{2D} \\
        & & \multicolumn{2}{c}{Truth Label} &&& \multicolumn{2}{c}{Truth Label} \\
        & & $N_{\pi^{0}} = 0$ & $N_{\pi^{0}} \geq 1$ &&& $N_{\pi^{0}} = 0$ & $N_{\pi^{0}} \geq 1$ \\
        \midrule
        \multirow{2}{*}{\STAB{\rotatebox[origin=c]{90}{Pred.}}}
        \multirow{2}{*}{\STAB{\rotatebox[origin=c]{90}{Label}}}
        & $N_{\pi^{0}} = 0$      & \bfseries 0.953 &  0.056            &&& \bfseries 0.953 &  0.097 \\
        & $N_{\pi^{0}} \geq 1$   & 0.047           &  \bfseries 0.944  &&& 0.047           &  0.903 \\
        \bottomrule
    \end{tabular}
    \label{tab:confusion_matrix_npi}
\end{table}

\subsection{Physics interpretation}

The physics reach of any large-scale LArTPC neutrino experiments is dictated by more subtle factors than just event classification accuracy. To illuminate the differences in ability to select certain categories of events, we use the classifications to emulate an analysis and selection for several important physics channels of relevance to long-baseline neutrino oscillation experiments, such as DUNE. The networks we trained predict a score for each category of event that can be interpreted as a probability.  For example, the neutrino classification score will predict, for each event, the probability for the event to be $\nu_e$ CC, $\nu_\mu$ CC, or NC, where the probabilities sum to 1.  We can create an event selection by applying thresholds to the probabilities of events we are targeting, for example $P_{\nu_e \,\text{CC}} > 90\%$ would select all events with at least 90\% probability to be electron neutrino charged current interactions.  We don't expect the 2D and 3D networks to have equal probability distributions even for the same collection of events, and so we optimize a figure of merit (FOM) for each analysis independently. The FOM is defined as the ratio of selected signal events to the square root of all selected events (signal + background).  The FOM is a function of the probability output by the networks, and for analyses targeting multiple branches of the network we optimize over the product of the probabilities.  In the physics analyses below, we present the best FOM for 2D compared to the best FOM for 3D.  In all cases, we also compare the selections for fixed values of both purity and efficiency. As another exercise, for each analysis we set the target purity (or efficiency) for an analysis to be equal in both 2D and 3D, and we compare the efficiency (or purity) at that requirement.

\subsubsection{Inclusive electron-neutrino charged-current interactions}

\begin{figure}
    \centering
    \includegraphics[width=.60\textwidth]{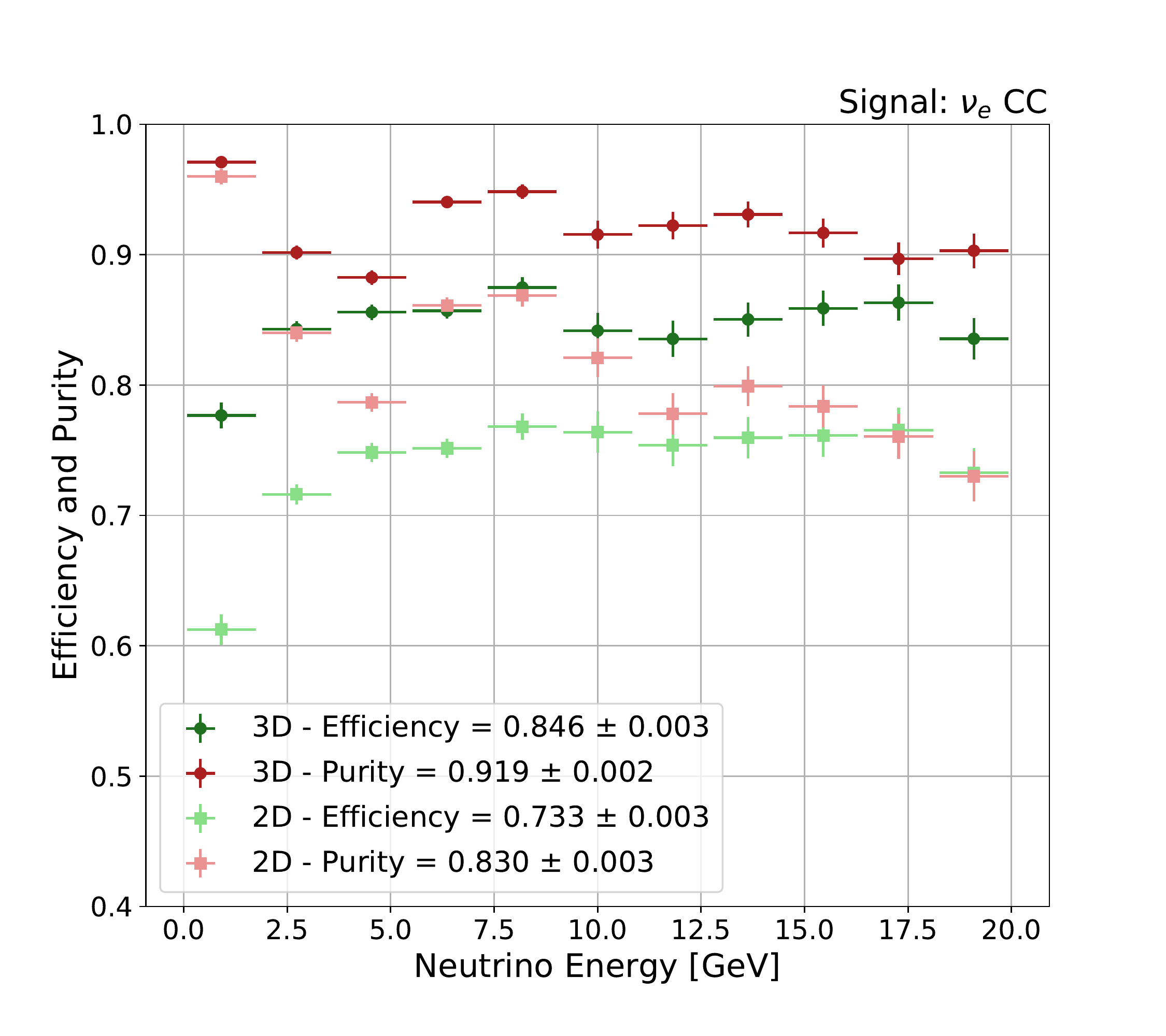}
    \caption{Efficiencies (green) and purities (red) as a function of neutrino energy for the inclusive $\nu_e$ CC selection. Results are shown for both 2D (light colors) and 3D (dark colors).}
    \label{fig:nue_cc_incl_eff_pur}
\end{figure}

Neutrino oscillation experiments require high efficiency to select electron-neutrino events with a low background contamination from neutral pion producing events which can emulate the topology of electron neutrino events and fake the signal. For this analysis we use the neutrino classification to select an inclusive electron-neutrino event classification and ignore the other categories. The results of this selection in terms of efficiencies and purities are shown in Figure~\ref{fig:nue_cc_incl_eff_pur}, the distributions of selected events are in Figure~\ref{fig:nue_cc_incl_evts}, the FOM and the ROC curves are available in Appendix~\ref{app:plots}.

\begin{figure}
    \centering
    \includegraphics[width=1.0\textwidth]{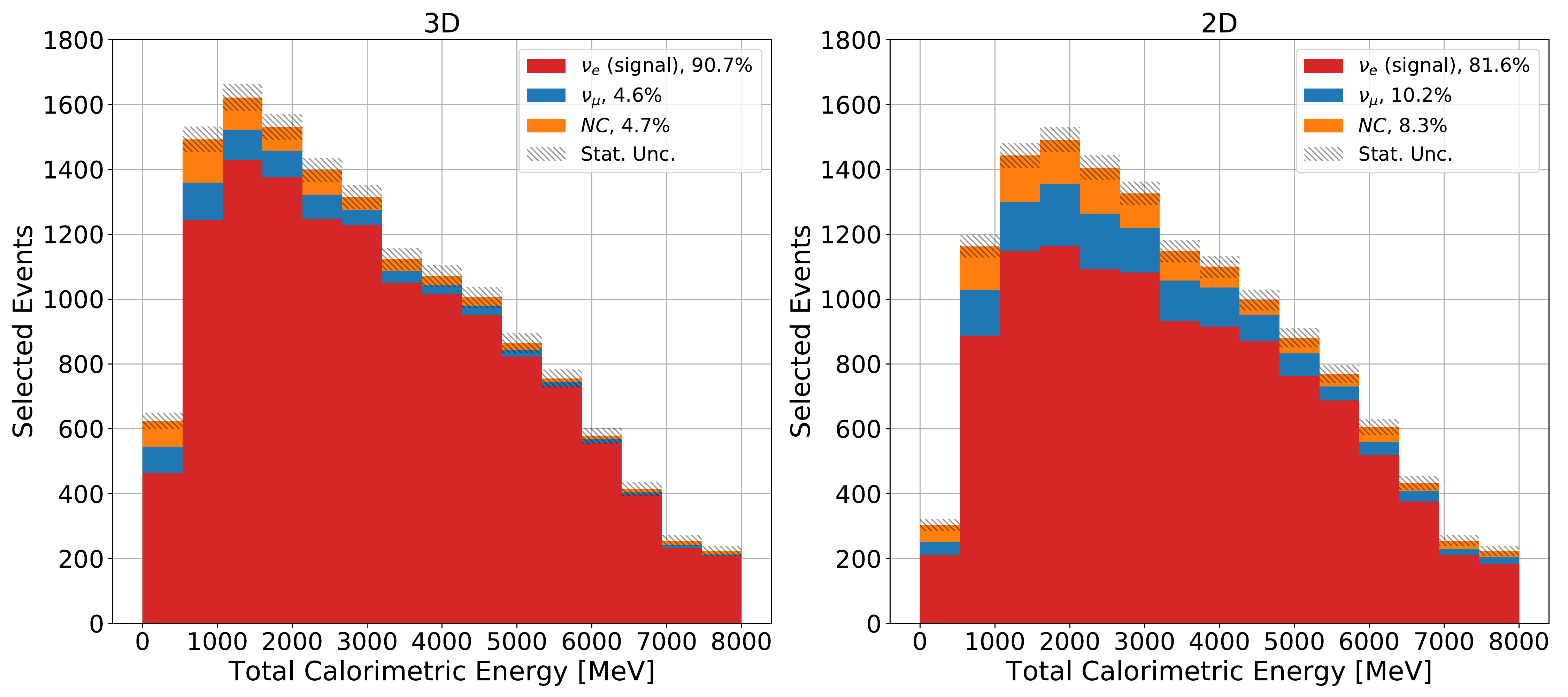}
    \caption{Selected events obtained with the inclusive $\nu_e$ CC selection as a function of the energy deposited for 3D (left) and 2D (right) selections. The horizontal axis is a measure of total \texttt{GEANT} energy depositions recorded in the TPC.}
    \label{fig:nue_cc_incl_evts}
\end{figure}


As an illustrative example for the inclusive $\nu_e $ CC analysis, we fix the efficiency for correct identification to be 85\% for both 2D and 3D readouts. With this selection the purity in the 2D projective readout is 62.7\% while in 3D is 90.6\%. This substantial gain in purity for the same efficiency is a huge advantage for the 3D pixel readout. 
Similarly if we instead fix the purity at 96\%, the efficiency is 34\% for 2D and 78\% for 3D. This demonstrates that with the same high-purity requirements, the 3D readout allows the selection of a significantly higher (more than double) number of neutrino interactions.

\subsubsection{Neutral current interactions with a neural pion}

\begin{figure}
    \centering
    \includegraphics[width=.60\textwidth]{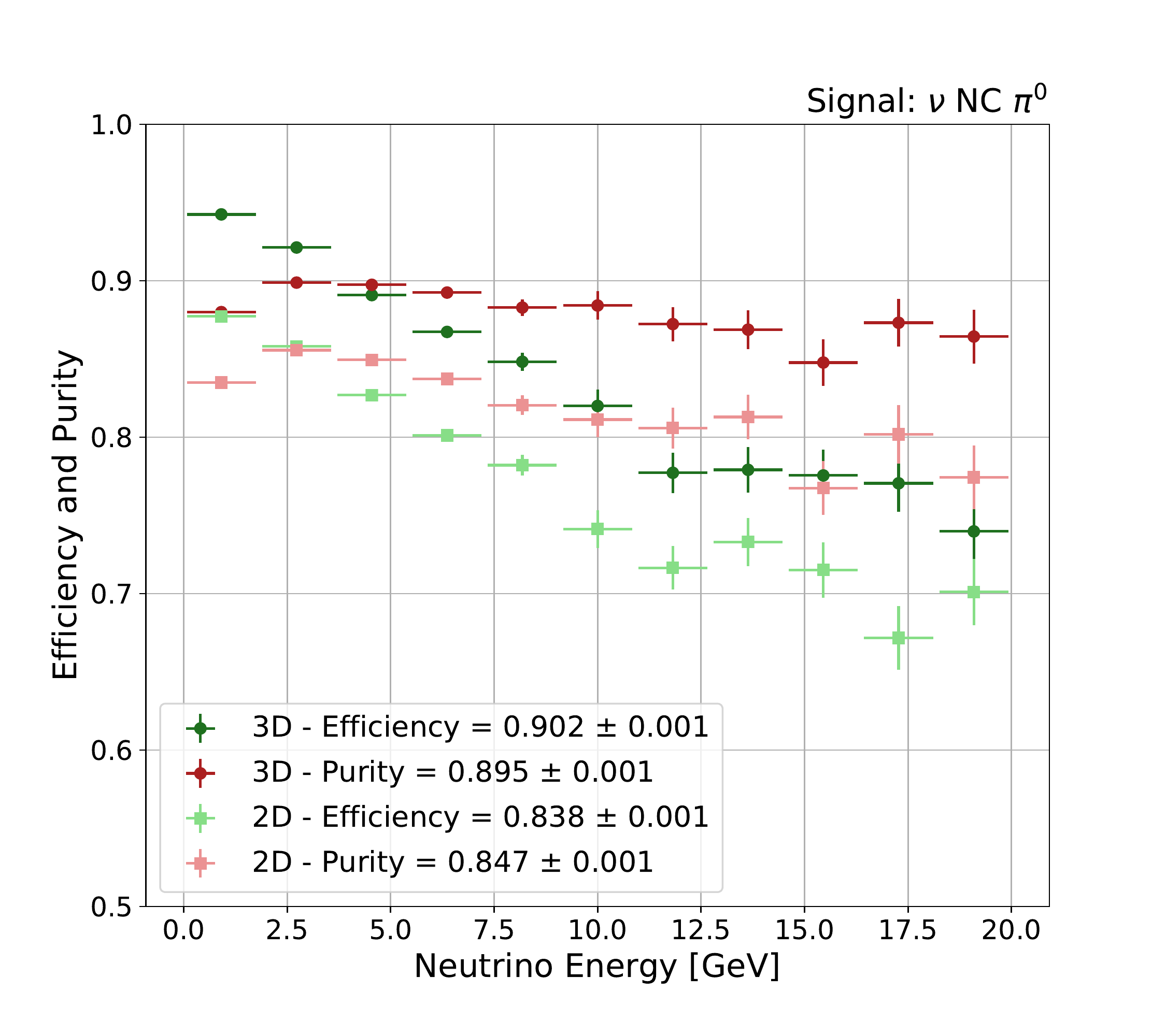}
    \caption{Efficiencies (green) and purities (red) as a function of neutrino energy for the $\nu$ NC $\pi^0$ selection. Results are shown for both 2D (light colors) and 3D (dark colors).}
    \label{fig:nu_nc_pi0_eff_pur}
\end{figure}

Neutral current events with a $\pi^0$ produced will form one of the most important backgrounds to the electron-neutrino search, as they also produce electromagnetic activity that could be mis-reconstructed.  The identification of neutral current events that produce a $\pi^0$ will be an essential analysis for constraining these backgrounds in the main electron-neutrino oscillation search. Therefore, we performed a selection that optimizes the same figure of merit, but in this case the signal is all events that are neutral current with a $\pi^0$ produced, and the background is any other events. 
The results of this selection are shown in Figure~\ref{fig:nu_nc_pi0_eff_pur}, while the distributions of selected events, the FOM and the ROC curves are available in Appendix~\ref{app:plots}. As in the electron-neutrino case, the selection efficiency and sample purity are better in 3D than in 2D.

As done in the previous subsection, we can fix the efficiency to be 85\% for both 2D and 3D and observe the final purity to be 83.5\% for 2D and of 93.1\% for 3D. Similarly, if we instead fix the purity at 98\%, the efficiency is 43\% for 2D and 63\% for 3D. Again this illustrates that with the same purity, the 3D readout allows to select a higher number ($\sim50\%$ improvement) of neutrinos.

\subsubsection{Inclusive muon-neutrino charged-current interactions}

\begin{figure}
\centering
    \includegraphics[width=.60\textwidth]{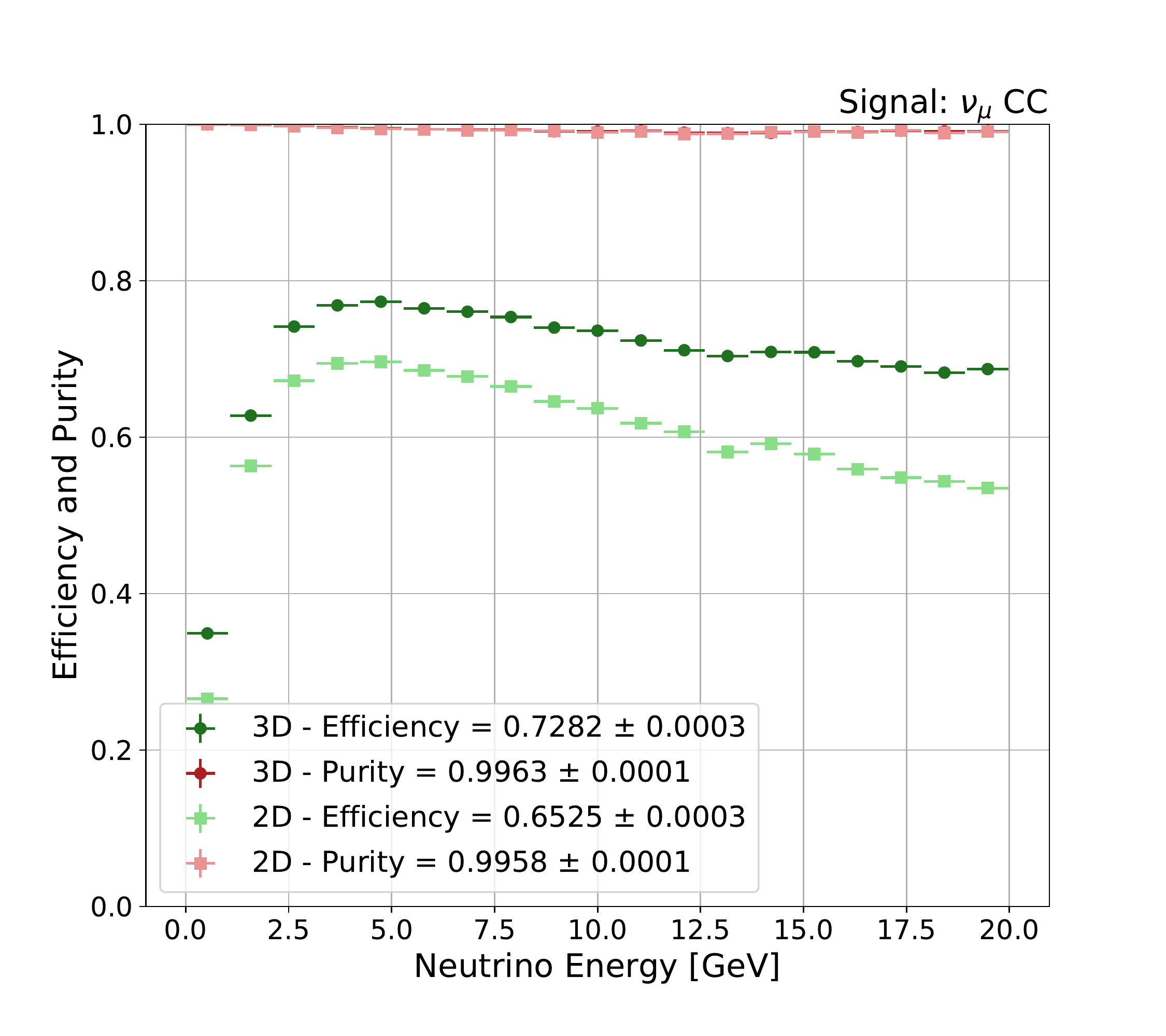}
    \caption{Efficiencies (green) and purities (red) as a function of neutrino energy for the inclusive $\nu_\mu$ CC selection. Results are shown for both 2D (light colors) and 3D (dark colors).}
    \label{fig:numu_cc_incl_eff_pur}
\end{figure}

Due to the use of a predominantly muon neutrino beam, the identification of $\nu_{\mu}$ CC events is a relatively easy selection to achieve high purity for both 2D and 3D. Therefore, rather than cutting to optimize signal significance as we have done previously, we set our cut such that we achieve 99\% background rejection and compare the measured efficiencies and purities for the 2D and 3D analyses. The results of this selection are shown in Figure~\ref{fig:numu_cc_incl_eff_pur}. The distributions of selected events, the FOM and the ROC curves are available in Appendix~\ref{app:plots}. As before, the muon neutrino selection has a higher efficiency for all neutrino energies in 3D, averaging 72\%, then in 2D, which averages 65\%.

\subsection{Future work}\label{Sec:Future}
For this initial, work we focused on a quantitative assessment of 2D projective readout against 3D pixel-based readout for neutrino interactions in a LArTPC using machine learning techniques. Machine learning techniques were chosen for this comparison to minimize the intrinsic performance differences specific reconstruction tools may have for the different types of readouts. In order to ensure a fair comparison of the two readouts, we assumed perfect detectors with the absence electronics response, noise, and the presence of non-responsive channels. Moreover, ambiguities in the 2D projective readout due to the particle trajectory parallel to specific readout channels were also ignored. 

Investigating the impact of noisy and dead channels on the reconstruction performances is a valuable avenue to pursue and is left for future studies. The amount of instrumented volume lost when a single channel dies is significantly higher in 2D versus 3D. However, the redundancy of 2D projection planes may partially shield the 2D projective detectors from the impact of dead channels.
The cumulative impact of these effects is expected to affect 2D wire-based projective readout more than 3D pixel-based one. This remains an important avenue to be study in order to quantify the resilience to such pathologies that may be afforded from an intrinsically 3D readout.

We also intend to quantify the performance of a 3D readout on the reconstruction of very-low-energy events in a LArTPC (compared to the DUNE energy spectrum used in this study). Neutrino interactions from supernova, the sun, the atmosphere or the Earth are an area of growing interest in the neutrino community. Additionally, we intend to study the ability of a 3D readout detector to identify and reconstruct rare events, such as those originating from proton decay and/or neutron-antineutron oscillations, which are signatures for evidence of physics beyond the Standard Model.

Finally, we intend to perform a campaign of hyperparameter optimization for the networks presented here, exploring both training parameters as well as sparse network designs. This work will require significant computational resources, and while we do not anticipate the optimal networks will significantly alter the conclusions of this paper, we do hope to guide the machine learning models used in future LArTPCs.

\section{Conclusions}\label{Sec:Conclusions}

In this work we made quantitative comparisons of the ability to identify neutrino events between a 2D projective wire-based readout and a 3D pixel-based readout for LArTPCs. We used the same set of neutrino interactions simulated from a high-energy, wide-band neutrino beam, such as that expected in the DUNE experiment. To enable as fair a comparison as possible, we leveraged the achievements in computer vision to train sparse, convolutional neural networks to perform multi-label event identifications.

The 3D pixel-based readout is found to be superior to the 2D projective one across a wide range of classifications. In particular, we found that for the identification of electron-neutrino events and the rejection of neutral current $\pi^0$ events, a 3D pixel-based detector significantly outperforms a 2D projective one by about a factor of two. Such a gain in efficiency and/or purity would directly translate into better physics performances or in lower running time for future experiments. While the exact quantification of this gain is experiment and analysis specific and must take into account a full suite of effects, the results here look very promising.

We have argued that, despite some assumptions and omissions to simplify the simulation of the readout technologies in both 2D and 3D, the conclusions of this study are unchanged by the assumptions. Moreover, the presence of wrapped wires currently envisioned in kiloton scale LArTPCs will only further complicate the identification and reconstruction of complex high-energy events. We do not model any such ambiguities, and instead assume the disambiguation due to wrapped wires to be perfect, yet still the 3D pixel-based readout performs better. Overall, when considering the unaccounted-for effects in our simulations, we believe that the stated capabilities of our best 2D networks are more overestimated for these detectors with respect to claims for 3D networks.

Future long-baseline neutrino experiments which will deploy kiloton-scale LArTPCs have the potential to make exciting new discoveries. Leveraging the technology to maximize its potential is a crucial task for the high energy physics community. The challenge to realize kiloton-scale pixel-based readout is a non-trivial one and efforts from the LArPix and Q-Pix groups are well underway. The pursuit of 3D pixel technology is well motivated by the foreseen physics impact presented here.

\section{Acknowledgements}\label{Sec:Acknowledgements}

 The authors gratefully thank B.J.P. Jones for fruitful early discussions. This research used resources of the Argonne Leadership Computing Facility, which is a DOE Office of Science User Facility supported under Contract DE-AC02-06CH11357. Researchers were supported by the Office of Science of the U.S. Department of Energy under award number DE-SC0020065.  This research used resources of the Oak Ridge Leadership Computing Facility at the Oak Ridge National Laboratory, which is supported by the Office of Science of the U.S. Department of Energy under Contract No. DE-AC05-00OR22725.









\bibliographystyle{JHEP}
\bibliography{bibliography}

\appendix

\section{Visualization of simulated events}

\begin{figure}
    \centering
    \includegraphics[width=0.3\linewidth]{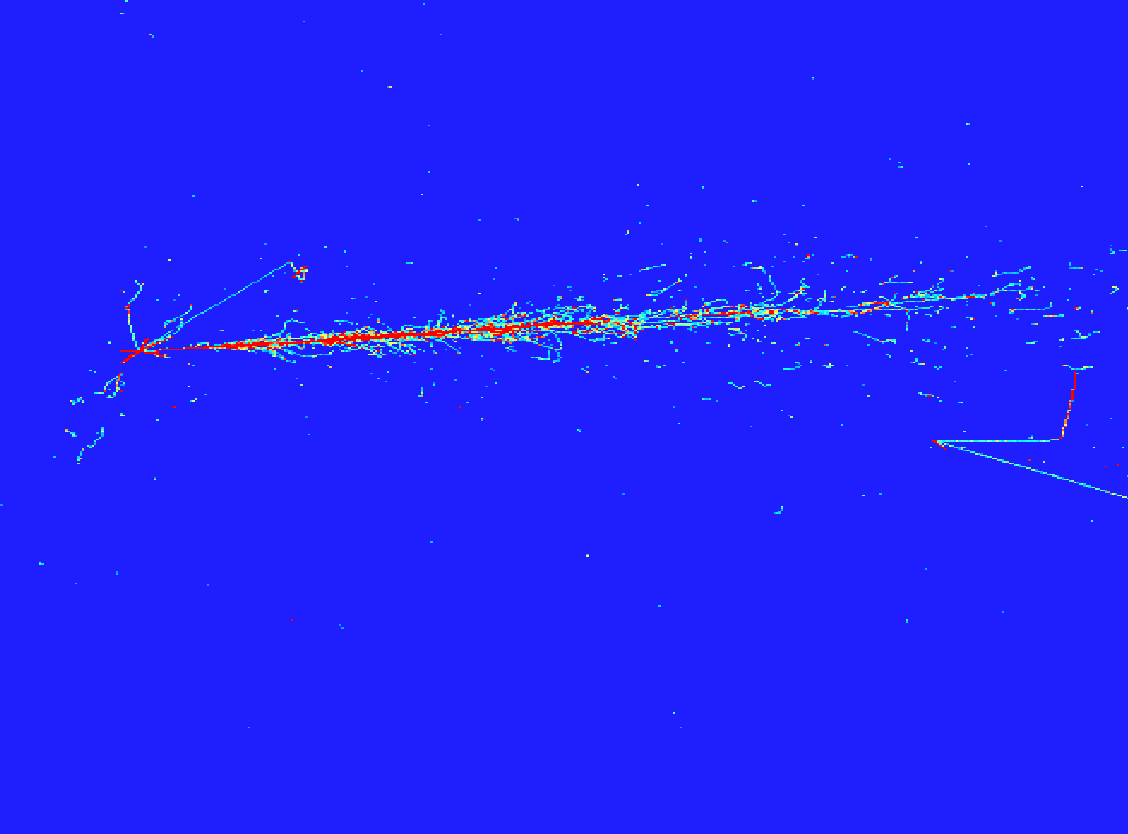}
    \includegraphics[width=0.3\linewidth]{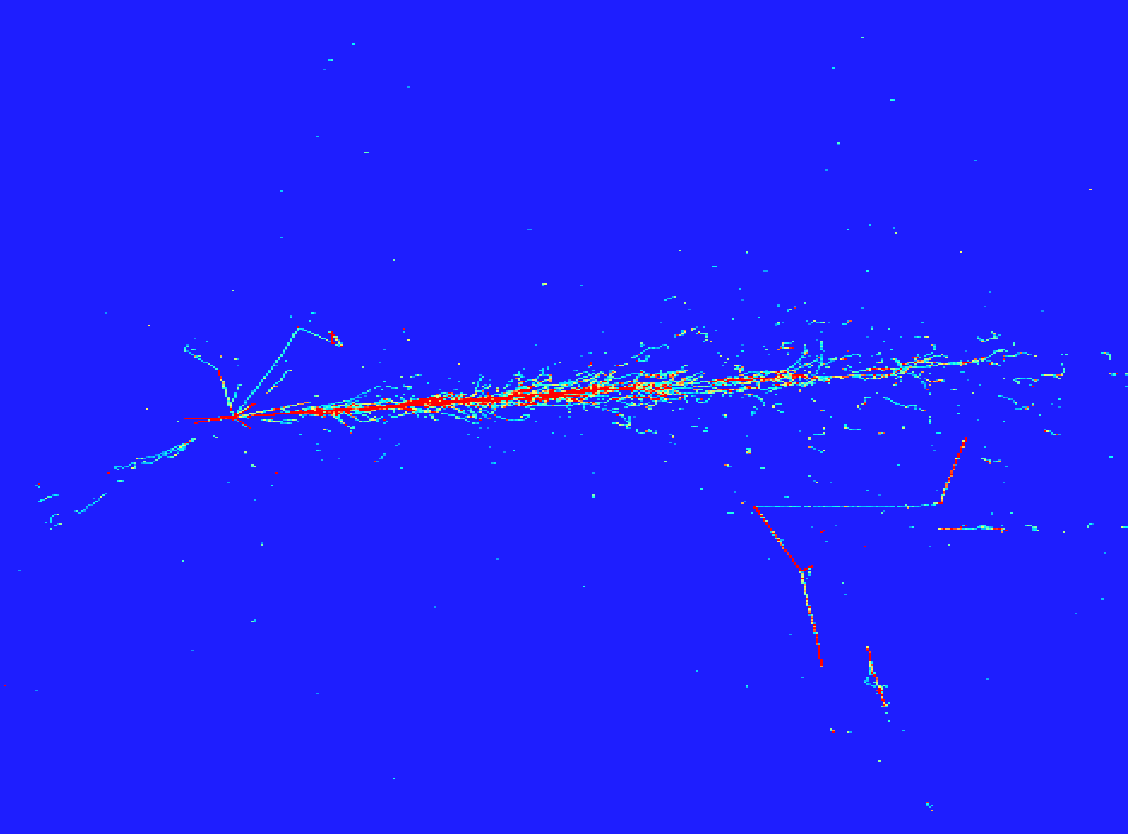}
    \includegraphics[width=0.3\linewidth]{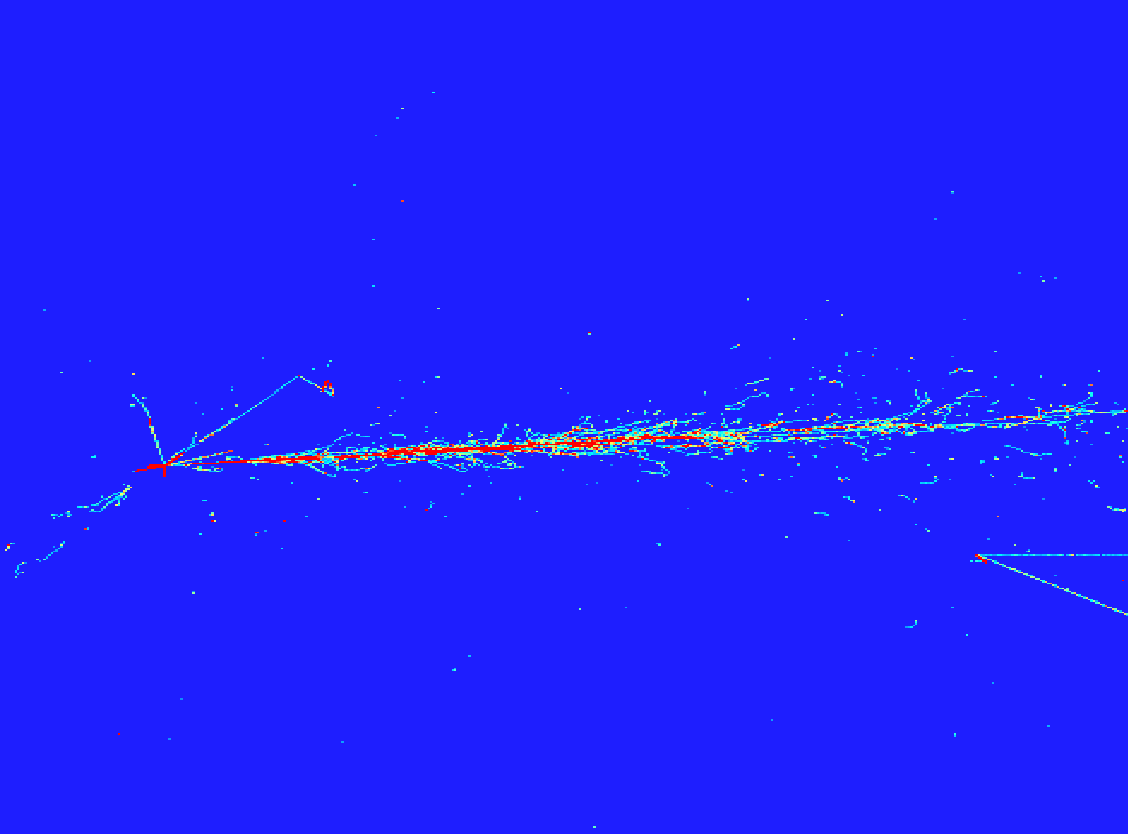}
    \includegraphics[angle=270, width=0.5\linewidth]{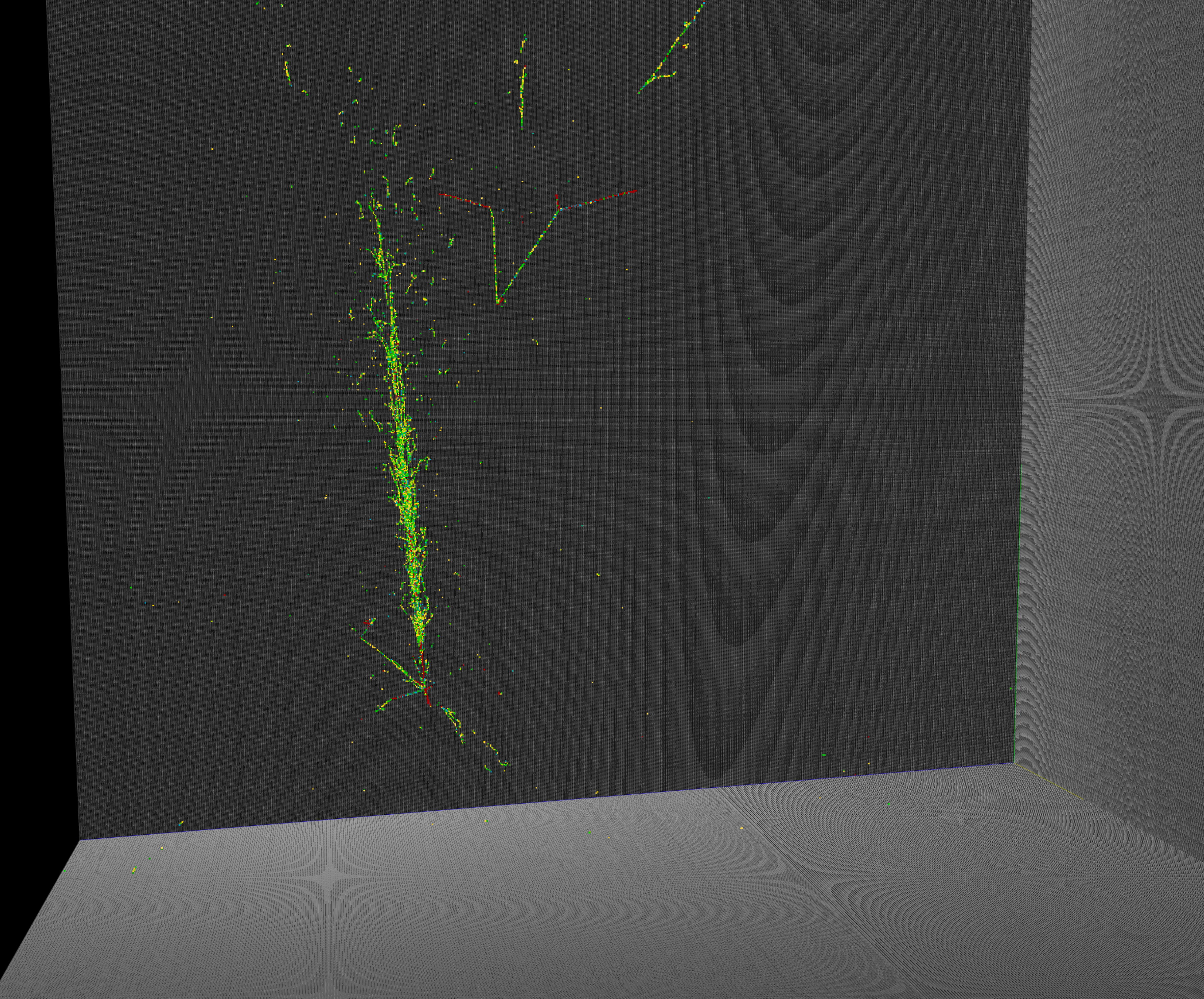}
    \caption{One electron-neutrino charged-current event as viewed in the three 2D projections (top) and in 3D (bottom).}
    \label{fig:evd1}
\end{figure}

\begin{figure}
    \centering
    \includegraphics[width=0.3\linewidth]{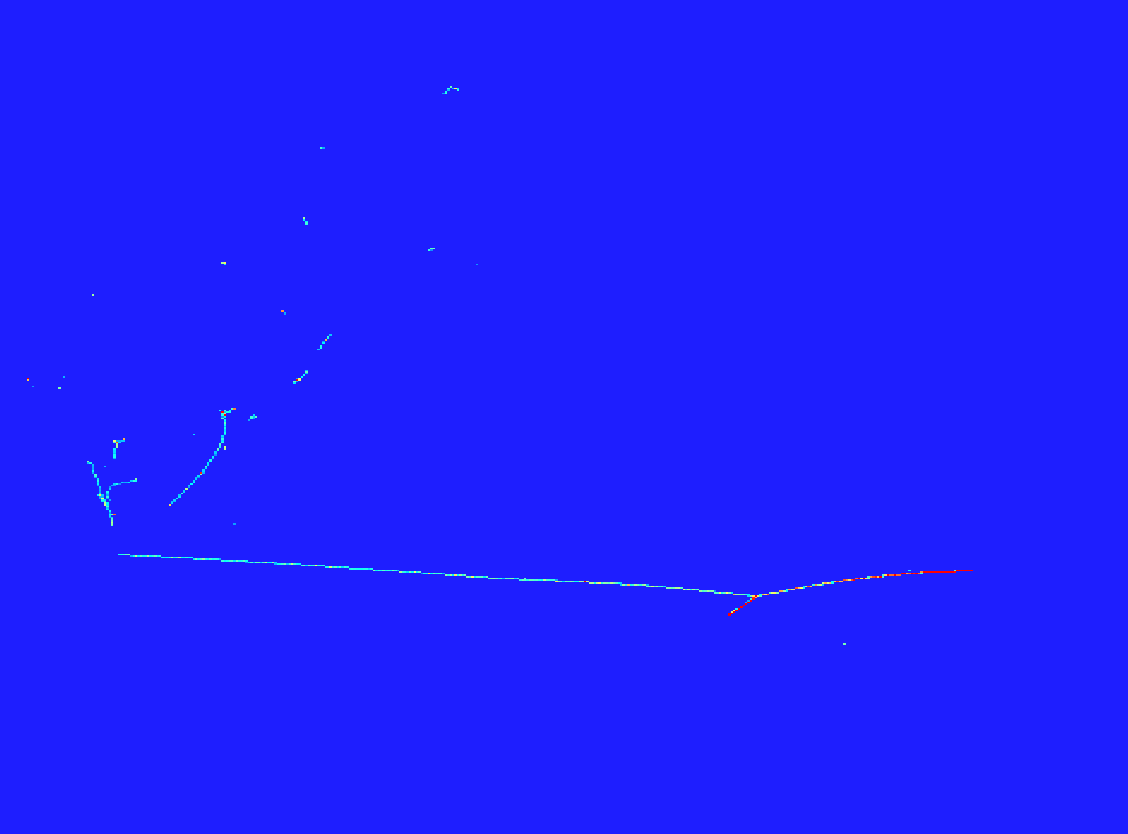}
    \includegraphics[width=0.3\linewidth]{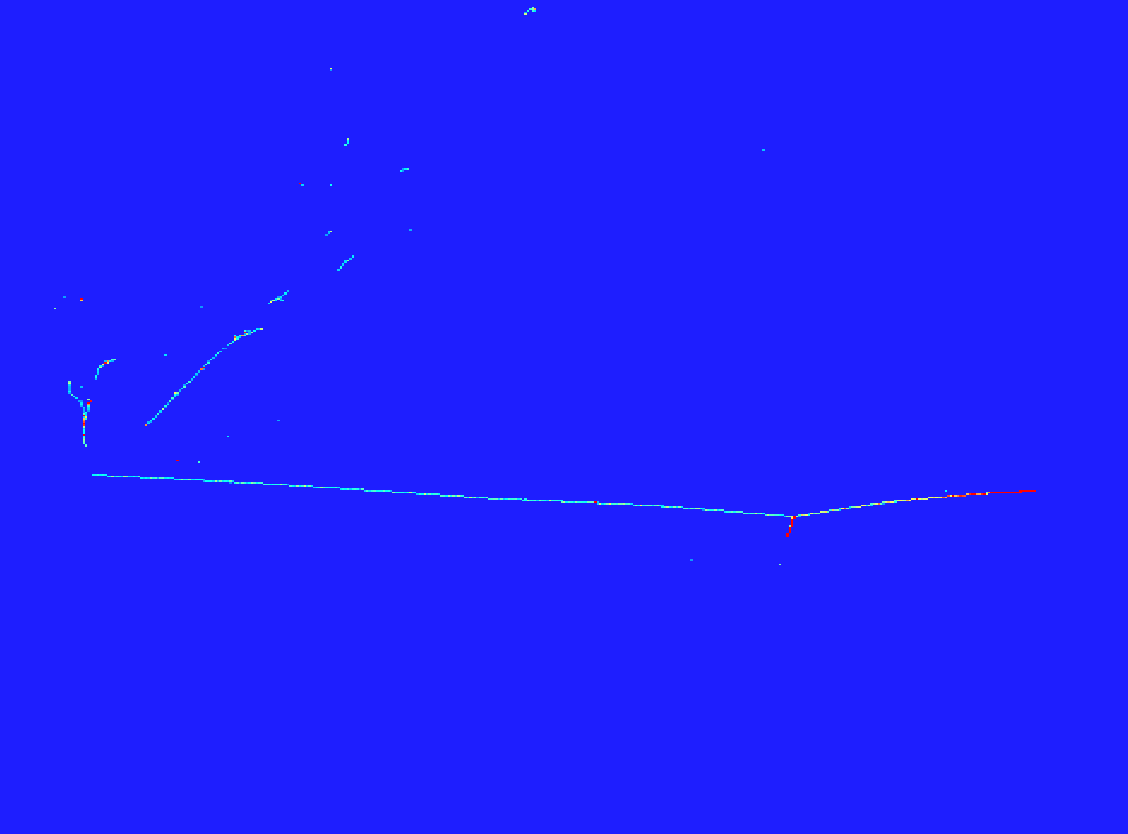}
    \includegraphics[width=0.3\linewidth]{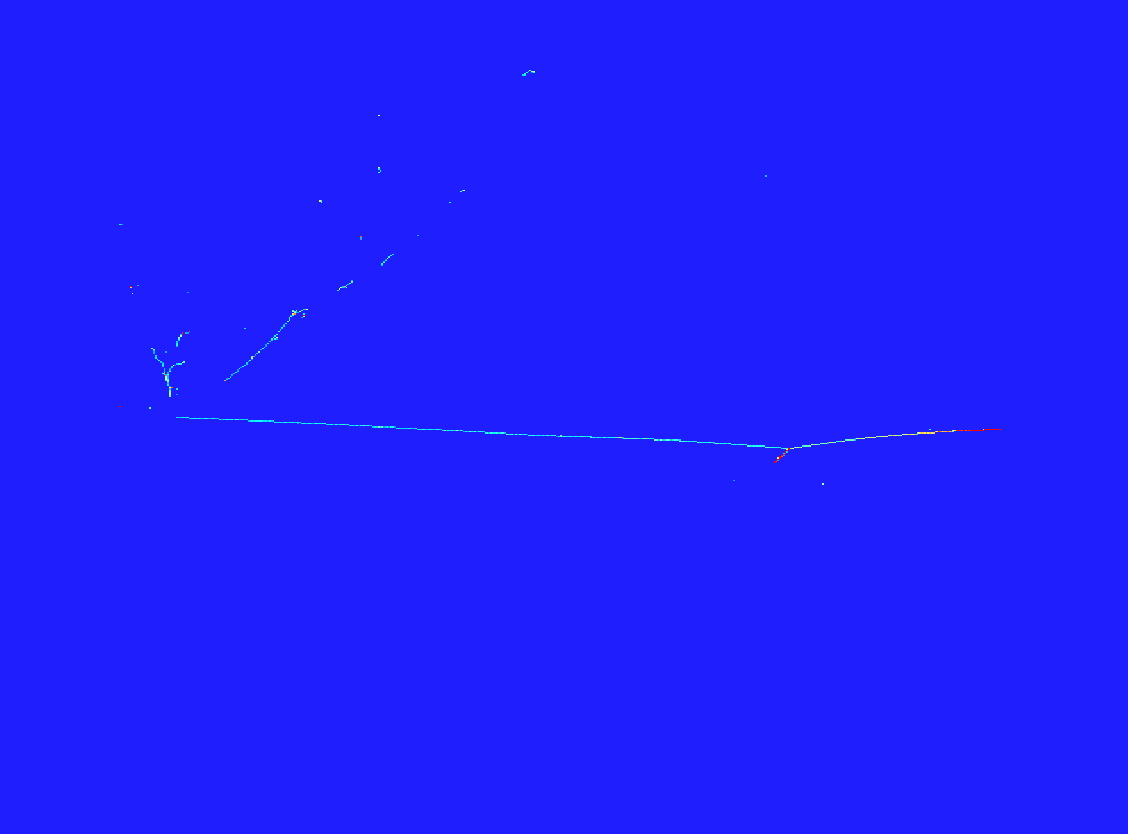}
    \includegraphics[angle=270, width=0.5\linewidth]{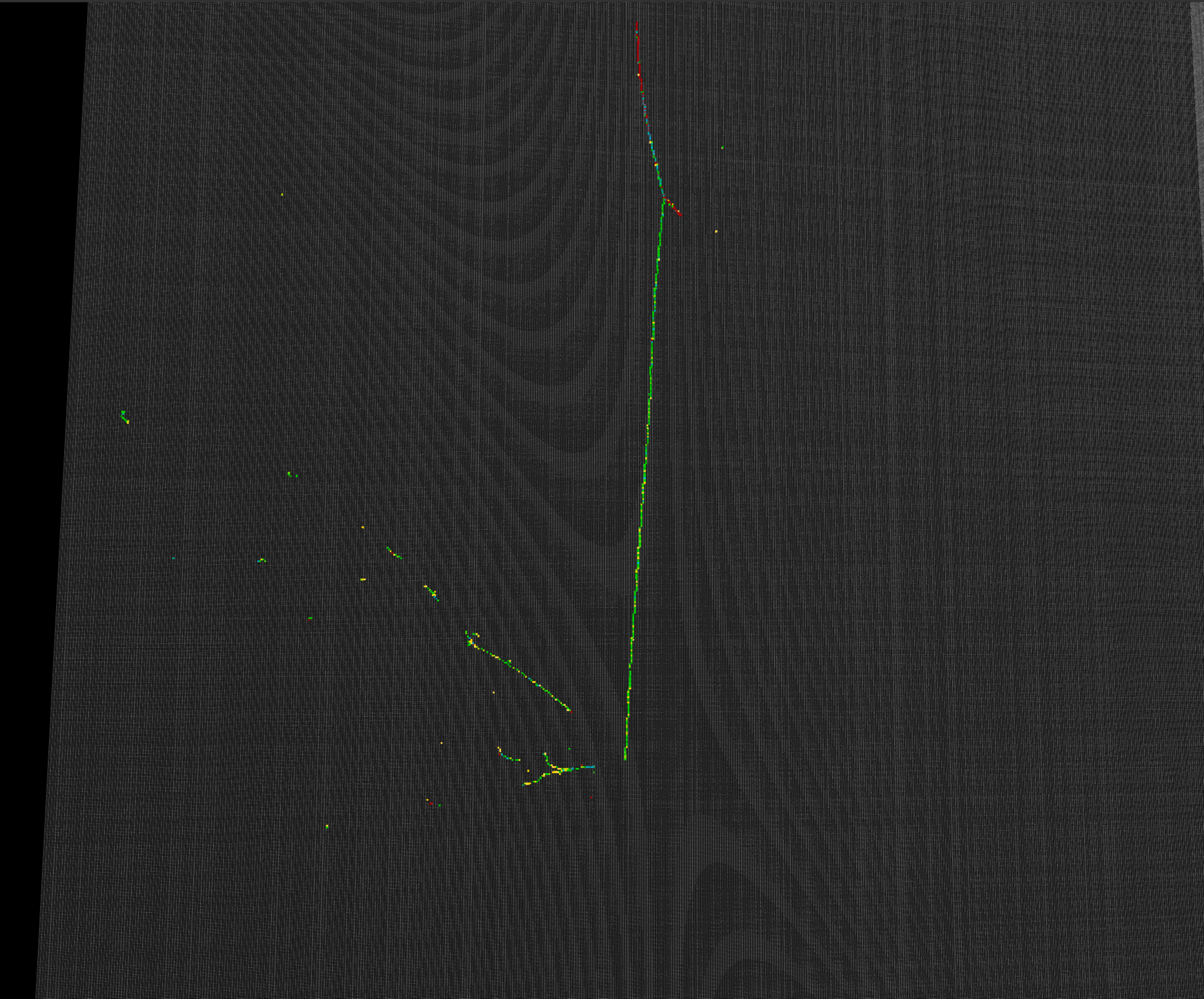}
    \caption{A neutral current event as viewed in the three 2D projections (top) and in 3D (bottom).}
    \label{fig:evd2}
\end{figure}

\begin{figure}
    \centering
    \includegraphics[width=0.3\linewidth]{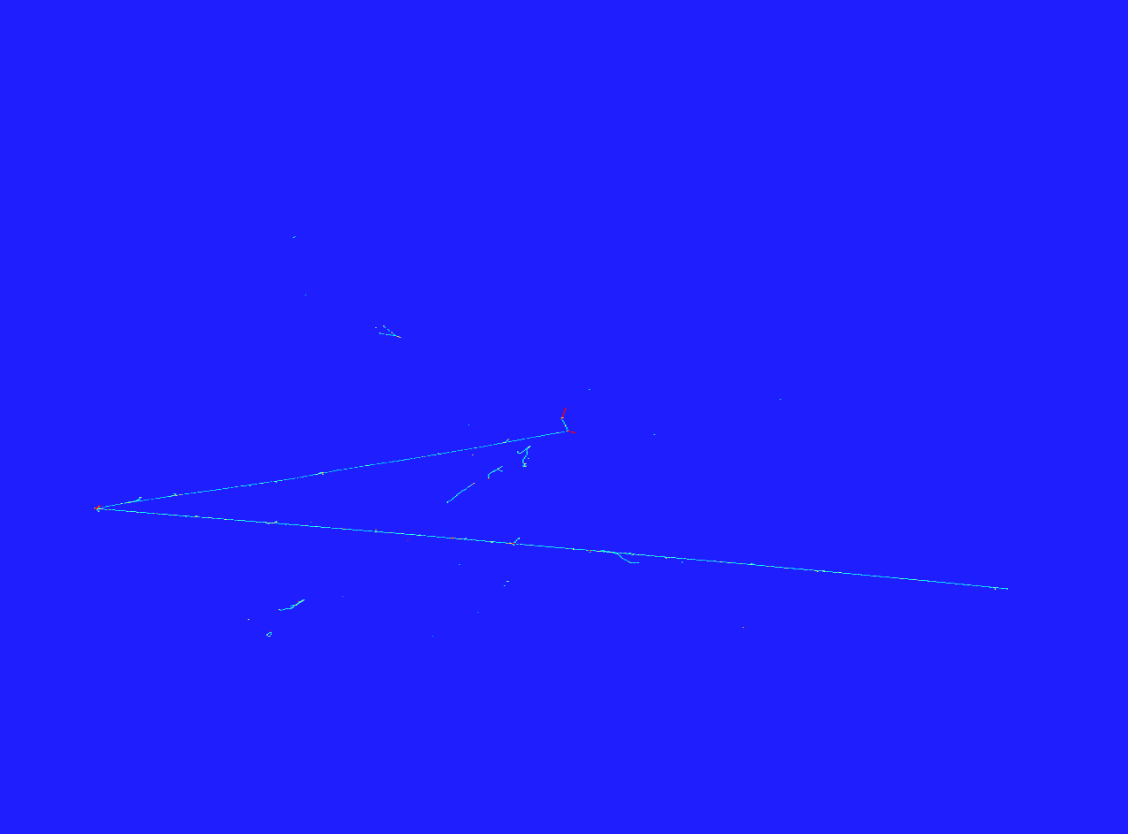}
    \includegraphics[width=0.3\linewidth]{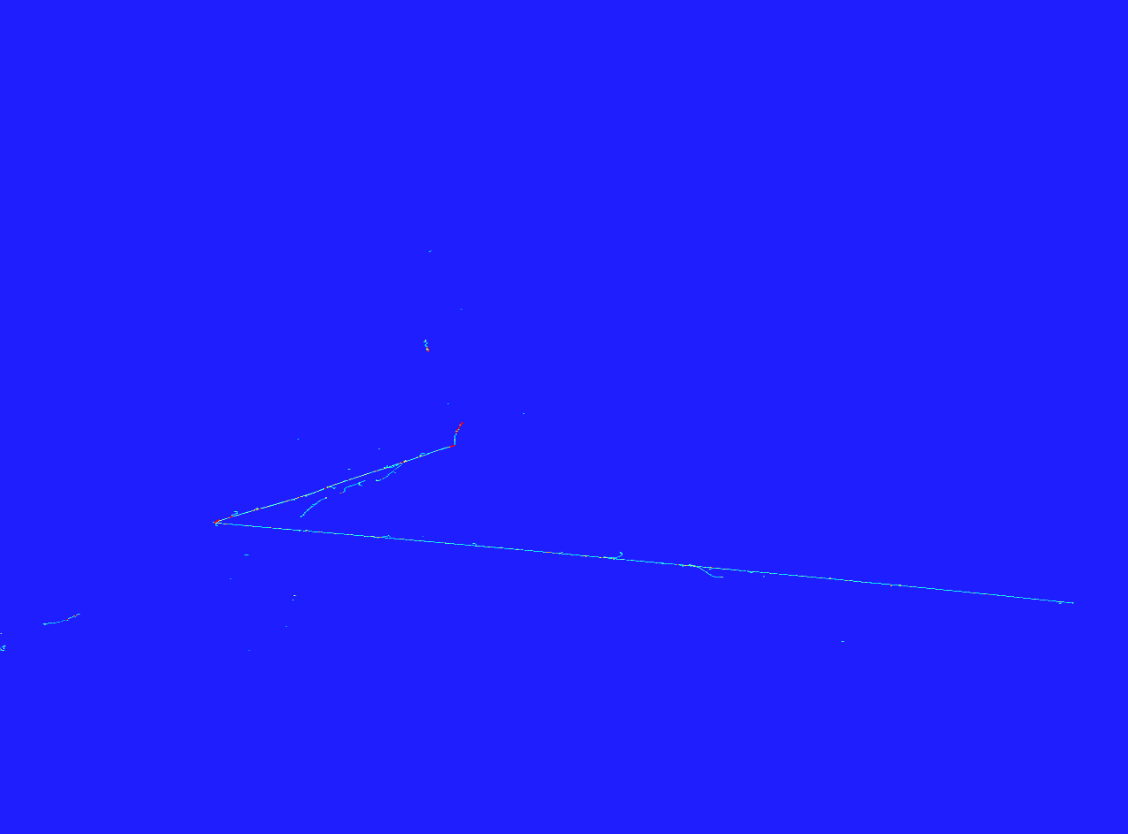}
    \includegraphics[width=0.3\linewidth]{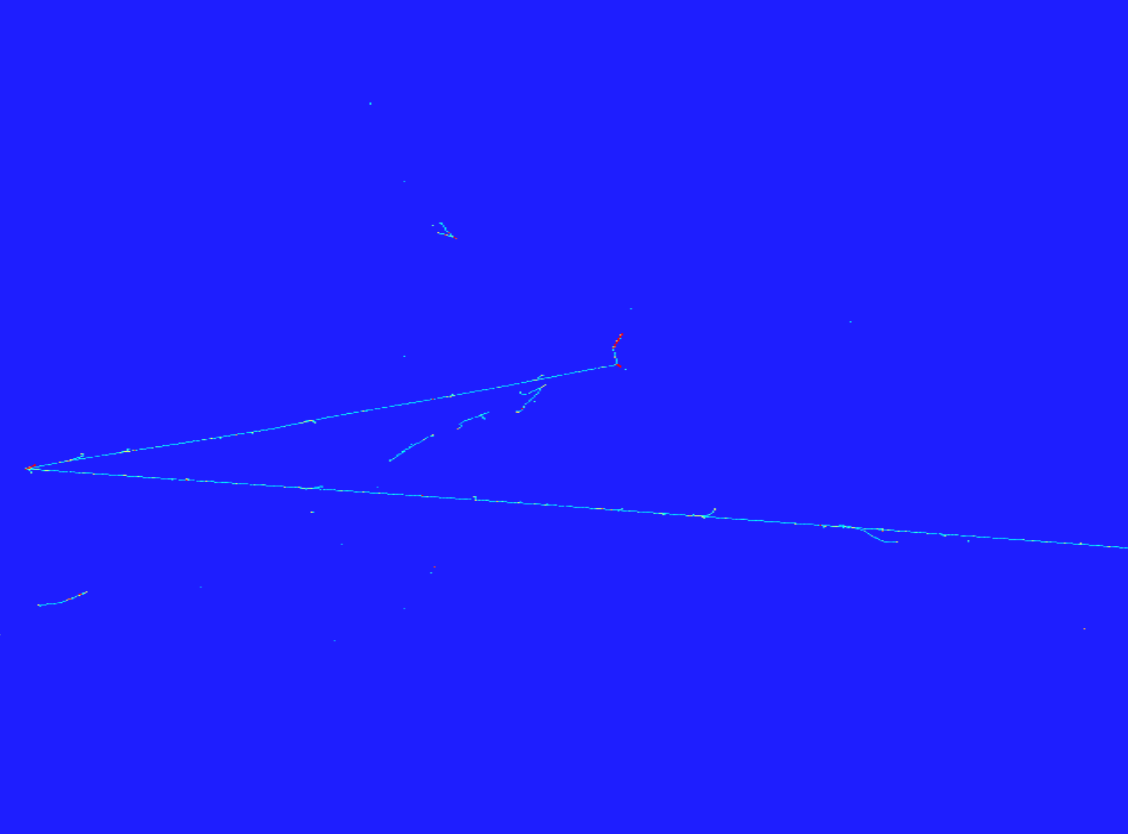}
    \includegraphics[angle=270, width=0.5\linewidth]{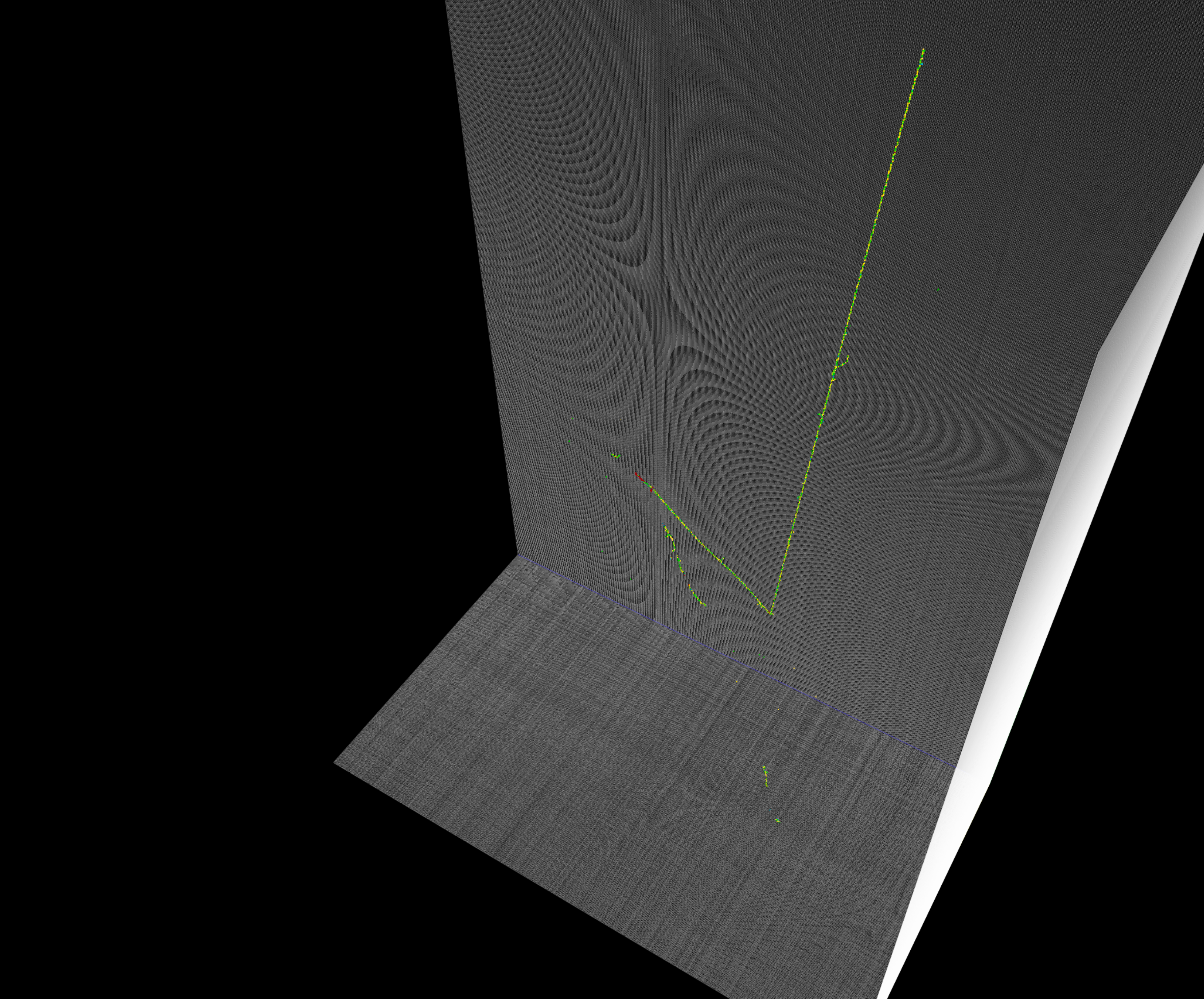}
    \caption{One muon-neutrino charged-current event as viewed in the three 2D projections (top) and in 3D (bottom).}
    \label{fig:evd3}
\end{figure}

In Figures~\ref{fig:evd1},~\ref{fig:evd2}, and~\ref{fig:evd3} we show more images of interactions for the 2D and 3D visualization of simulated events.

\section{Additional analysis plots}
\label{app:plots}

\begin{figure}
\centering
    \begin{subfigure}[]{0.8\textwidth}
        \includegraphics[width=\textwidth]{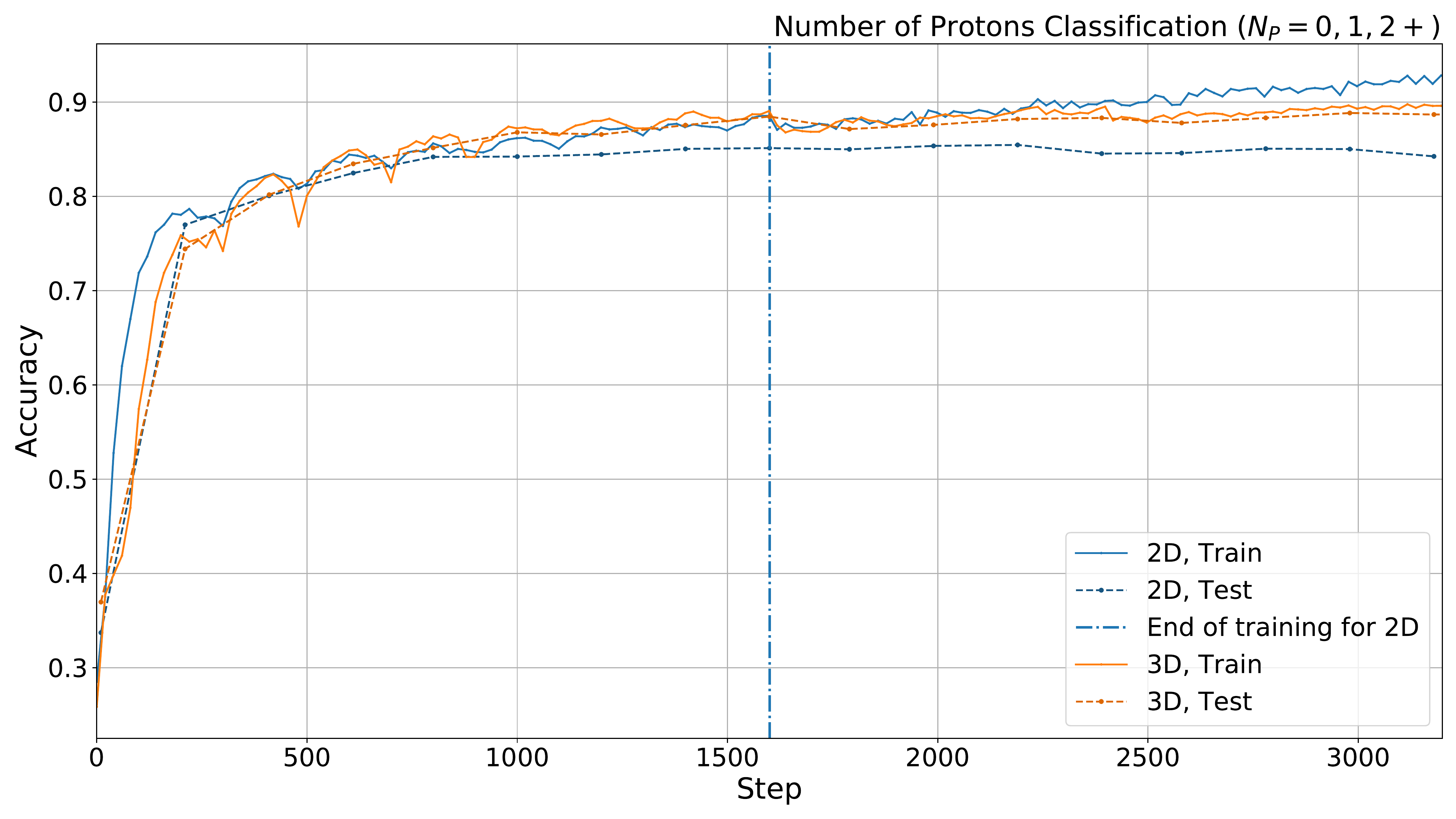}
        \caption{}
        \label{fig:accuracy_prot}
    \end{subfigure}
    \begin{subfigure}[]{0.8\textwidth}
        \includegraphics[width=\textwidth]{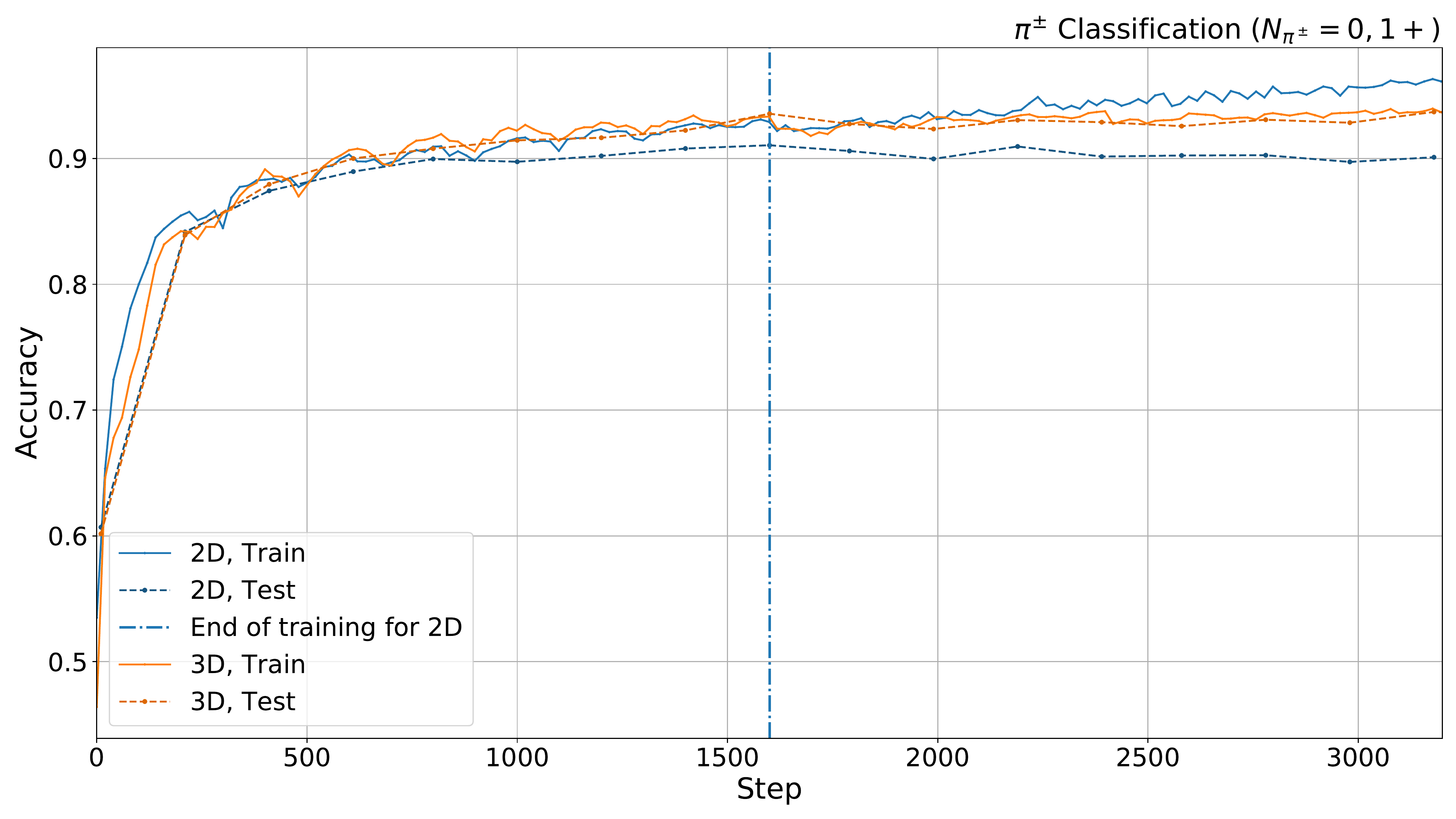}
        \caption{}
        \label{fig:accuracy_pi+}
    \end{subfigure}
    \begin{subfigure}[]{0.8\textwidth}
        \includegraphics[width=\textwidth]{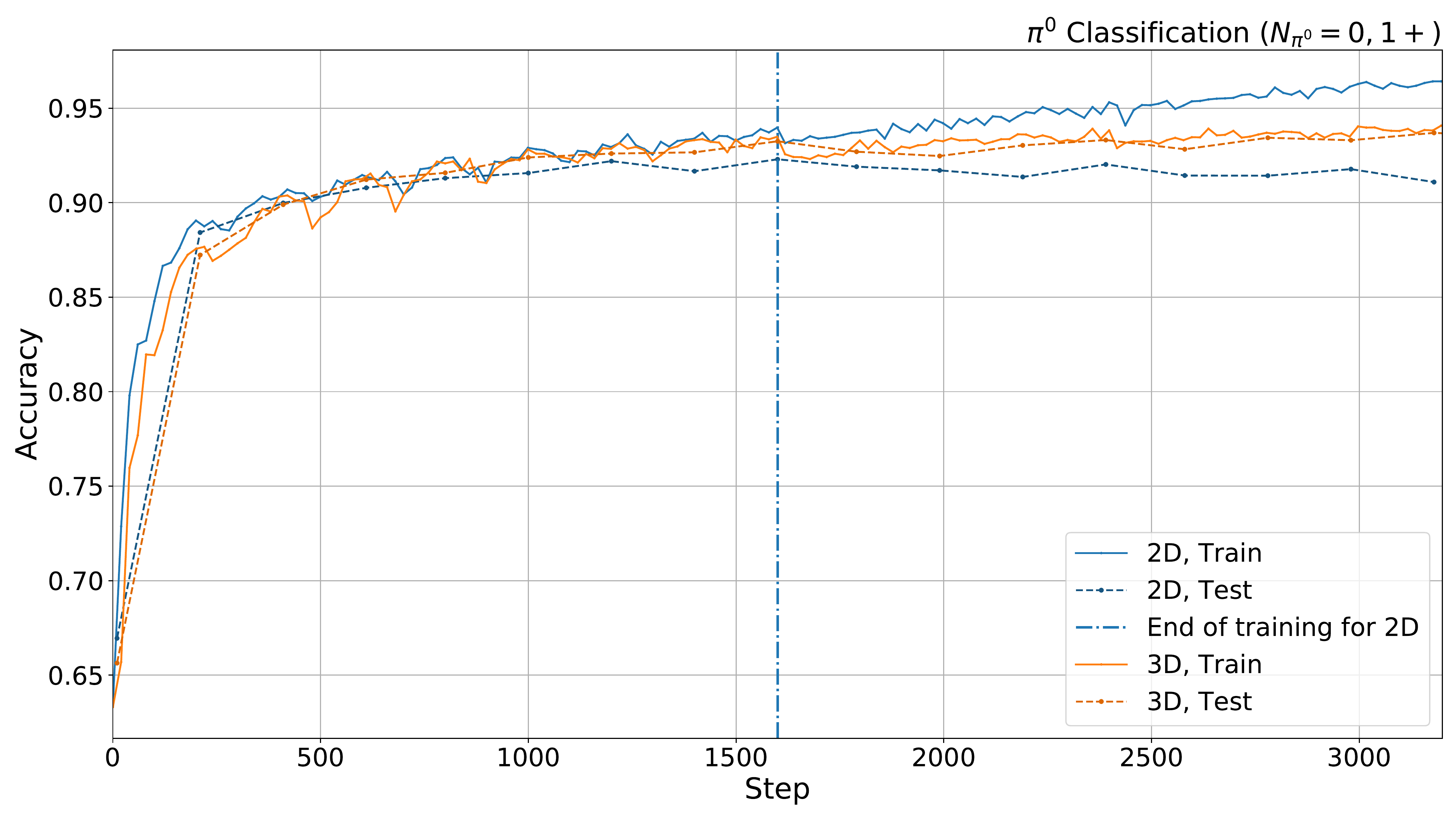}
        \caption{}
        \label{fig:accuracy_pi0}
    \end{subfigure}
    \caption{
        Training and testing accuracies for the proton (\subref{fig:accuracy_prot}), the charged pion (\subref{fig:accuracy_pi+}) and the neutral pion (\subref{fig:accuracy_pi0}) classifications for both the 2D (blue curves) and 3D (orange curves) networks. Although the 2D curves are shown up to iteration step 3200, for the following studies we pick the trained model as obtained at iteration 1600 (vertical dashed blue line), before the network begins overfitting.
    }
\label{fig:accuracy_loss_app}
\end{figure}

\begin{figure}
\centering
    \begin{subfigure}[]{0.495\textwidth}
        \includegraphics[width=\textwidth]{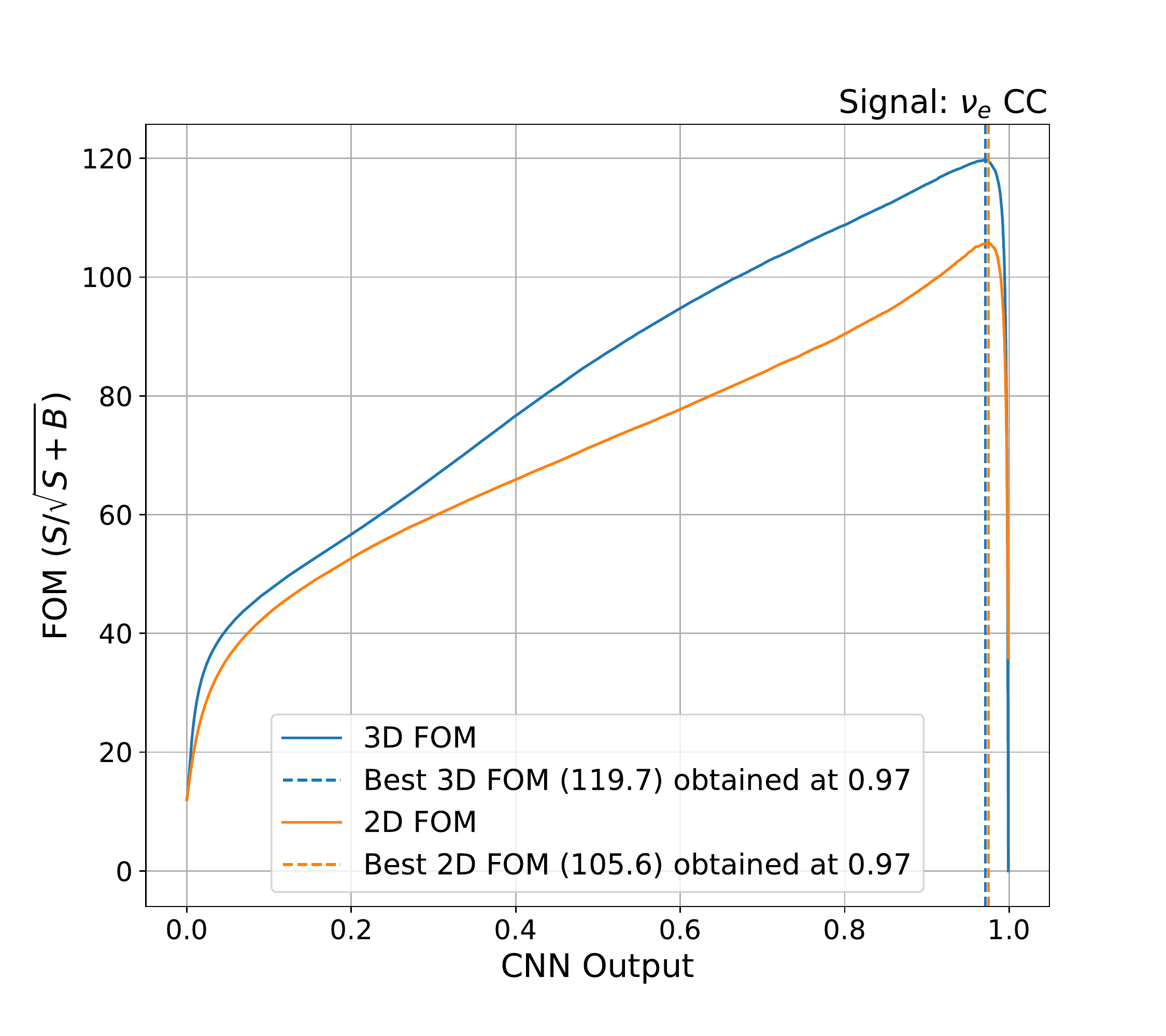}
        \caption{}
        \label{fig:nue_cc_incl_fom}
    \end{subfigure}
    \begin{subfigure}[]{0.495\textwidth}
        \includegraphics[width=\textwidth]{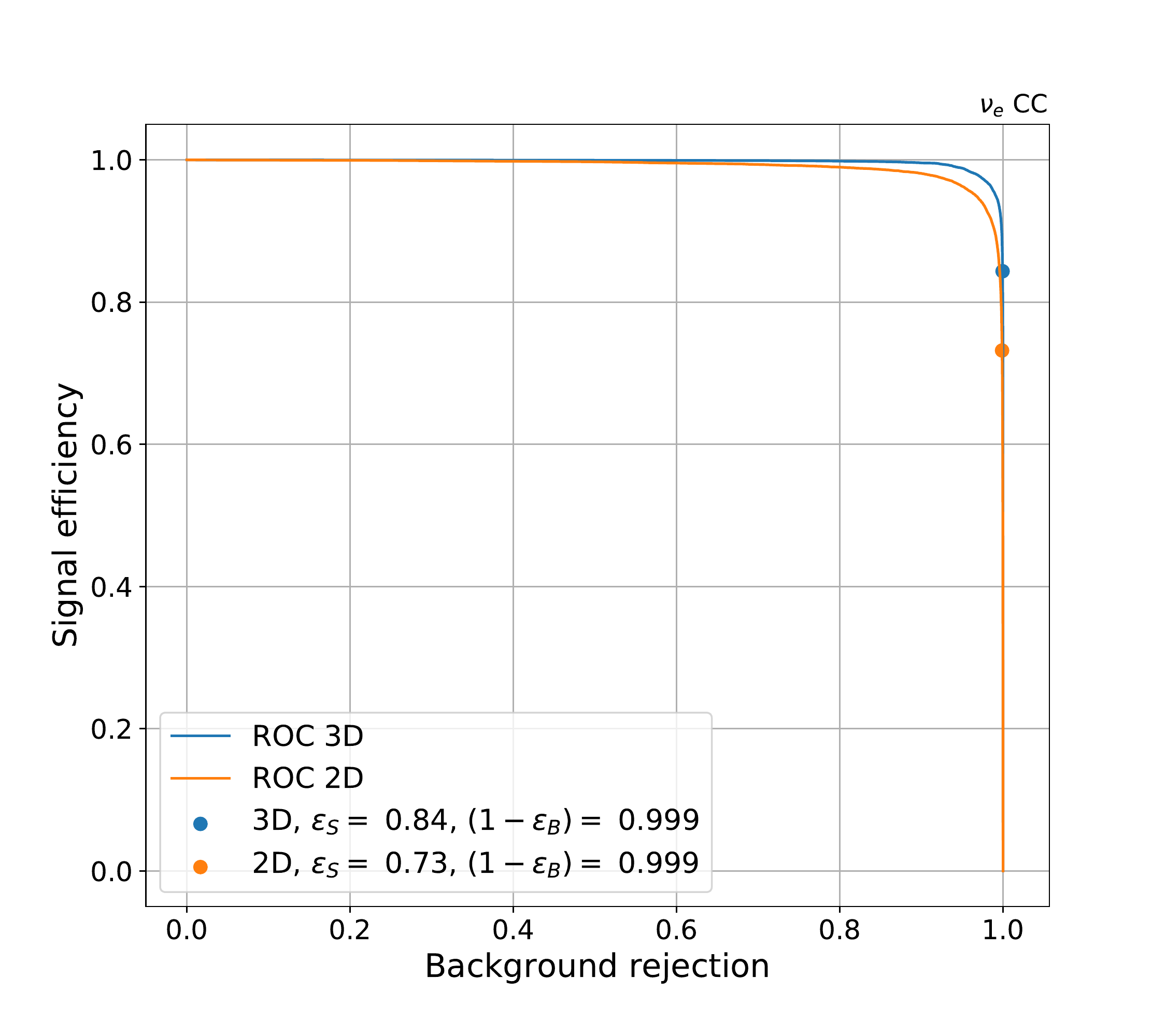}
        \caption{}
        \label{fig:nue_cc_incl_roc}
    \end{subfigure}
    \caption{
        \subref{fig:nue_cc_incl_fom} Figure of merit for the inclusive $\nu_e$ CC selection as a function of the cut applied on the CNN output for both the 2D and the 3D models. 
        \subref{fig:nue_cc_incl_roc} ROC curve showing the signal efficiency and the background rejection for different values of the cut applied on the CNN output. The two points correspond to the best cuts obtained maximizing the figure of merit. 
    }
\label{fig:nue_cc_incl_fom_roc}
\end{figure}

\begin{figure}
\centering
    \begin{subfigure}[]{0.495\textwidth}
        \includegraphics[width=\textwidth]{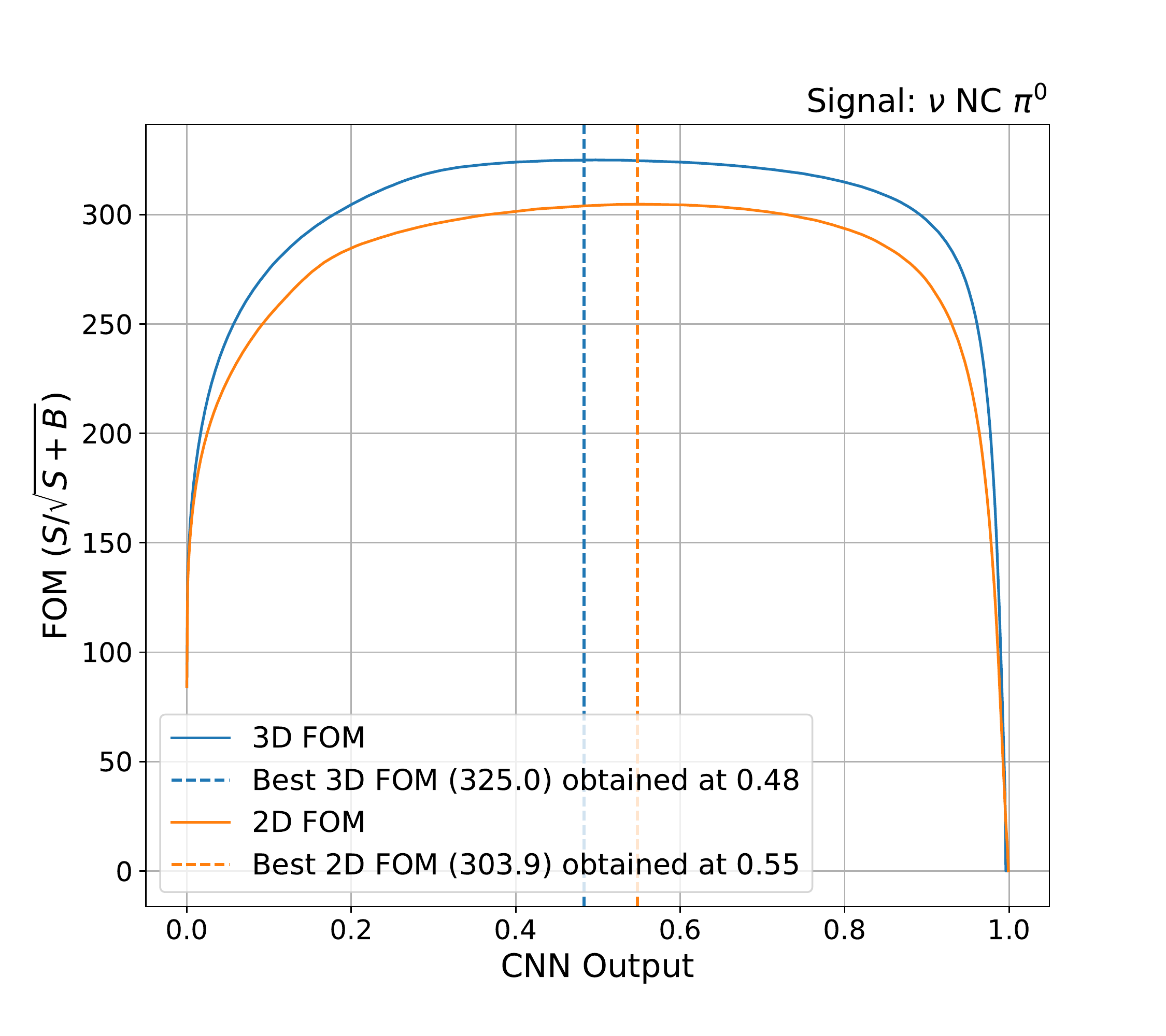}
        \caption{}
        \label{fig:nu_nc_pi0_fom}
    \end{subfigure}
    \begin{subfigure}[]{0.495\textwidth}
        \includegraphics[width=\textwidth]{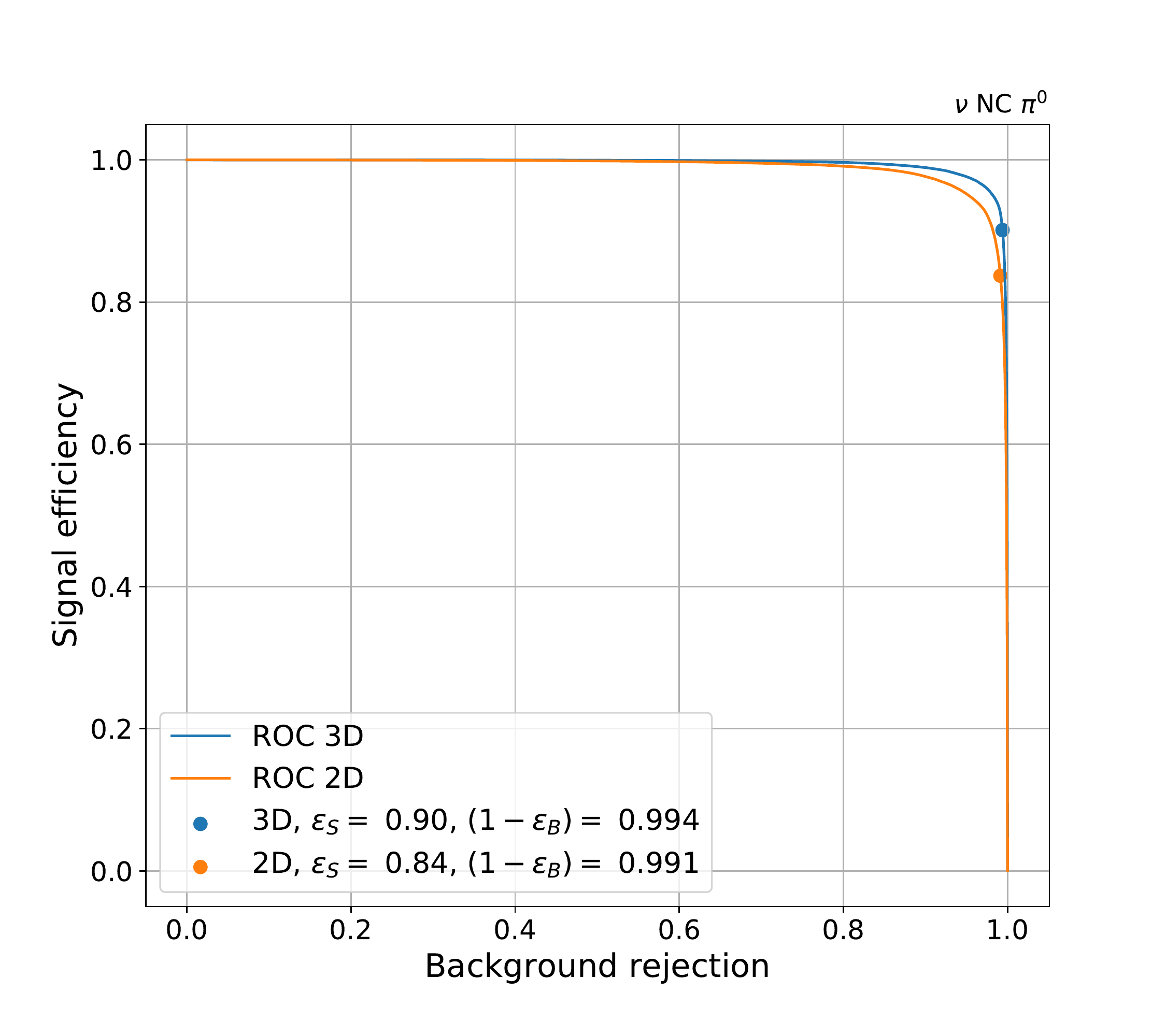}
        \caption{}
        \label{fig:nu_nc_pi0_roc}
    \end{subfigure}
    \caption{
        \subref{fig:nu_nc_pi0_fom} Figure of merit for the $\nu$ NC $\pi^0$ selection as a function of the cut applied on the CNN output for both the 2D and the 3D models. 
        \subref{fig:nu_nc_pi0_roc} ROC curve showing the signal efficiency and the background rejection for different values of the cut applied on the CNN output. The two points correspond to the best cut obtained maximizing the figure of merit. 
    }
\label{fig:nu_nc_pi0_roc_fom}
\end{figure}

\begin{figure}
    \centering
    \includegraphics[width=1.0\textwidth]{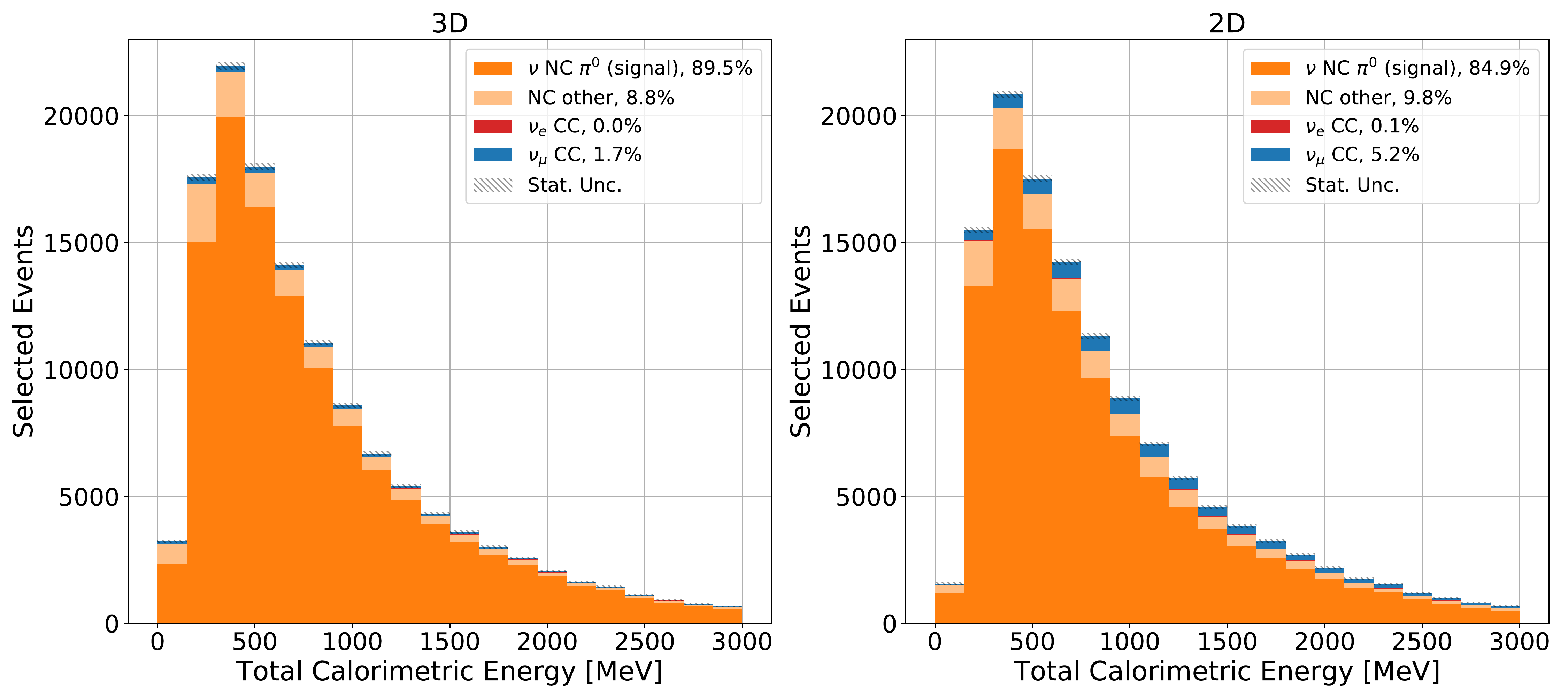}
    \caption{Selected events obtained with the $\nu$ NC $\pi^0$ selection as a function of the energy deposited for 3D (left) and 2D (right) selections. The horizontal axis is a measure of total \texttt{GEANT} energy depositions recorded in the TPC.}
    \label{fig:nu_nc_pi0_evts}
\end{figure}

\begin{figure}
    \centering
    \begin{subfigure}[]{0.495\textwidth}
        \includegraphics[width=\textwidth]{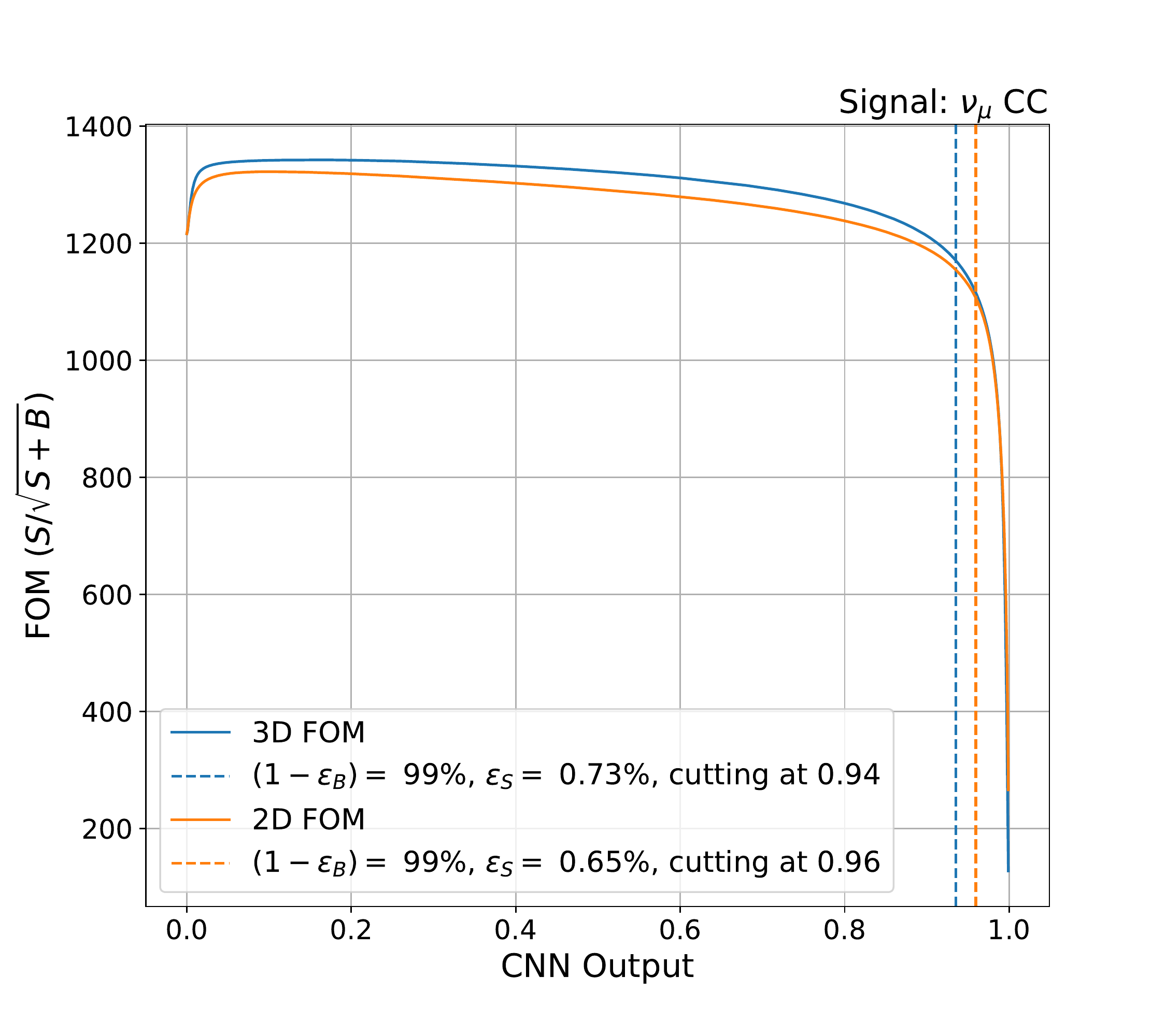}
        \caption{}
        \label{fig:numu_cc_incl_fom}
    \end{subfigure}
    \begin{subfigure}[]{0.495\textwidth}
        \includegraphics[width=\textwidth]{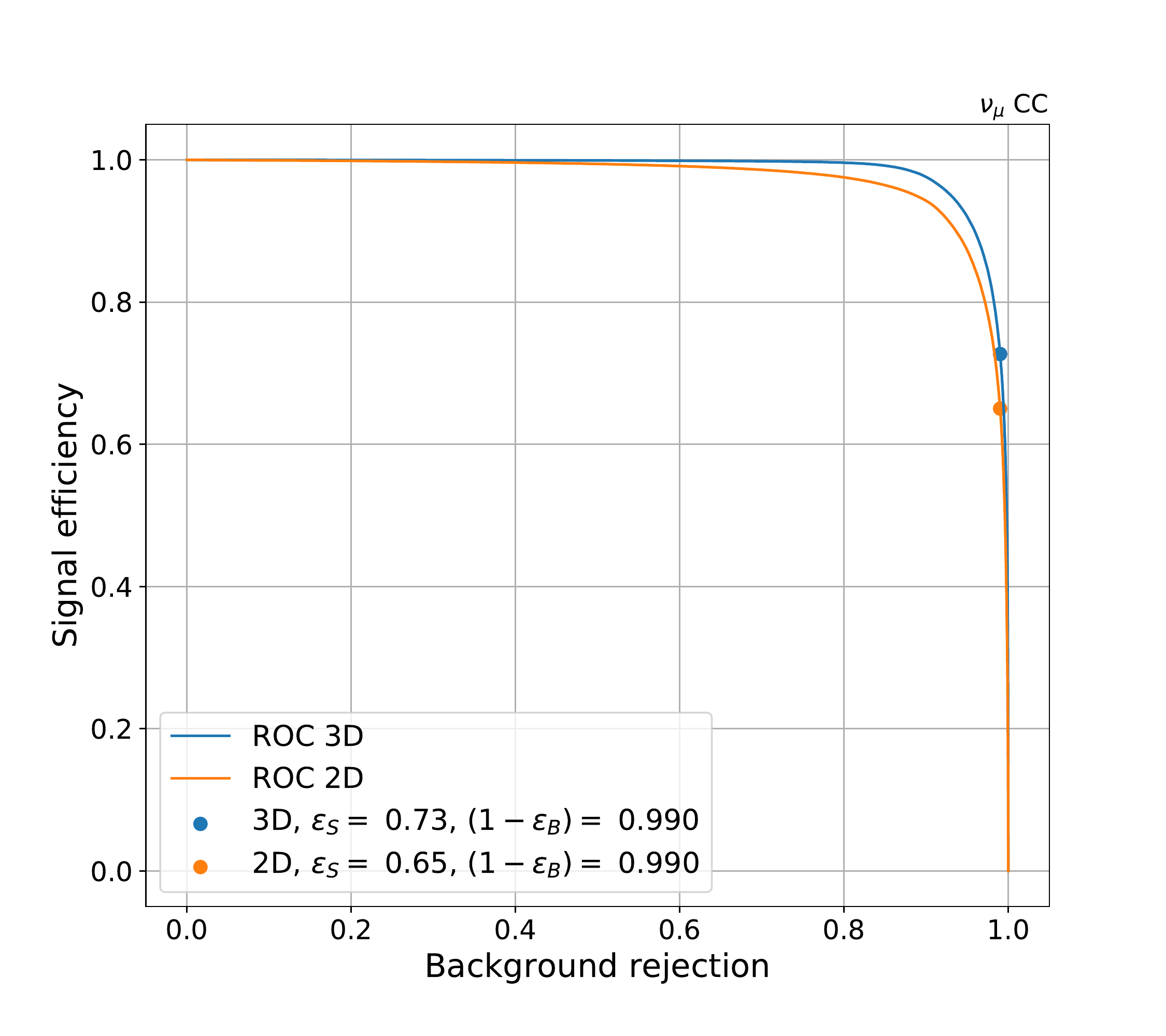}
        \caption{}
        \label{fig:numu_cc_incl_roc}
    \end{subfigure}
    \caption{
        \subref{fig:numu_cc_incl_fom} Figure of merit for the inclusive $\nu_\mu$ CC selection as a function of the cut applied on the CNN output for both the 2D and the 3D models. 
        \subref{fig:numu_cc_incl_roc} ROC curve showing the signal efficiency and the background rejection for different values of the cut applied on the CNN output. The two points correspond to the best cut obtained maximizing the figure of merit. 
    }
\label{fig:numu_cc_incl_roc_fom}
\end{figure}

\begin{figure}
    \centering
    \includegraphics[width=1.0\textwidth]{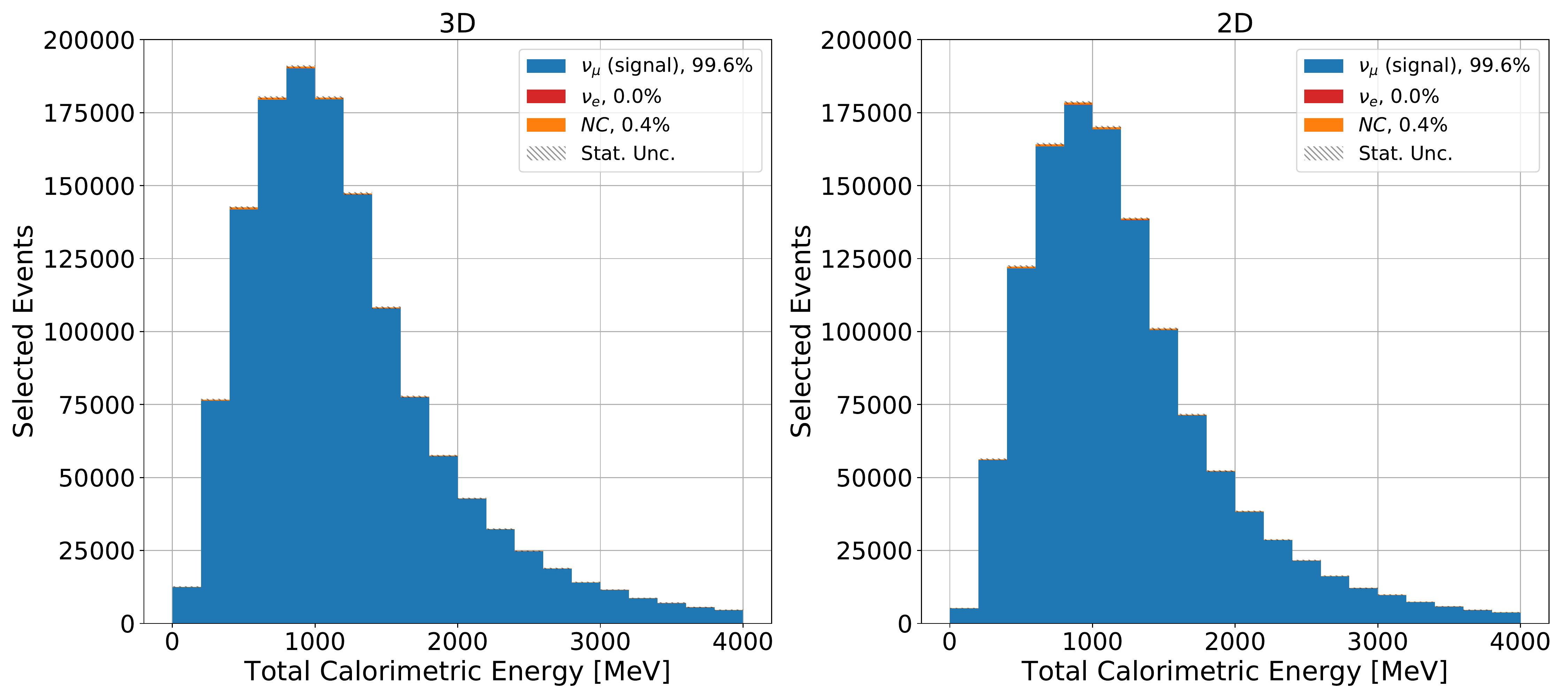}
    \caption{Selected events obtained with the inclusive $\nu_\mu$ CC selection as a function of the energy deposited for 3D (left) and 2D (right) selections. The selection cut applied to the CNN output is optimized so that both analyses have a background rejection of 99\%, which gives a purity of 99.6\%. The horizontal axis is a measure of total \texttt{GEANT} energy depositions recorded in the TPC.}
    \label{fig:numu_cc_incl_evts}
\end{figure}

In this appendix we show additional plots related to the trained network and the analyses presented above. 
Figure~\ref{fig:accuracy_loss_app} shows the training and testing accuracies for the proton multiplicity, the presence of charged pion and the presence of neutral pion classifications.
Figure~\ref{fig:nue_cc_incl_fom_roc} shows the FOM and the ROC curve for the inclusive $\nu_e$ CC analysis.
Figures~\ref{fig:nu_nc_pi0_roc_fom} and~\ref{fig:nu_nc_pi0_evts} show the FOM, the ROC curve and the selected events distributions for the $\nu$ NC $\pi^0$ analysis.
Finally Figures~\ref{fig:numu_cc_incl_roc_fom} and ~\ref{fig:numu_cc_incl_evts} show the FOM, the ROC curve and the selected events distributions for the inclusive $\nu_\mu$ CC analysis.

\end{document}